\documentclass[pre,twocolumn,longbibliography]{revtex4-1}
\usepackage[cp850]{inputenc}
\usepackage{natbib}
\usepackage{amssymb,amsmath}
\usepackage{epsfig}
\usepackage{bm}
\usepackage{color}

\usepackage{graphicx}

\usepackage{color}

\newcommand\beq{\begin{equation}}
\newcommand\eeq{\end{equation}}
\newcommand\beqa{\begin{eqnarray}}
\newcommand\eeqa{\end{eqnarray}}
\newcommand{\dd}{\text{d}}

\newcommand{\al}{\alpha}

\begin{document}

\title{Enskog kinetic theory for multicomponent granular suspensions}

\author{Rub\'en G\'omez Gonz\'alez\footnote[1]{Electronic address: ruben@unex.es}}
\affiliation{Departamento de F\'{\i}sica,
Universidad de Extremadura, E-06006 Badajoz, Spain}
\author{Nagi Khalil\footnote[2]{Electronic address: nagi.khalil@urjc.es}}
\affiliation{Escuela Superior de Ciencias Experimentales y Tecnolog\'{\i}a (ESCET), Universidad Rey Juan Carlos, M\'ostoles 28933, Madrid, Spain}
\author{Vicente Garz\'{o}\footnote[3]{Electronic address: vicenteg@unex.es;
URL: http://www.unex.es/eweb/fisteor/vicente/}}
\affiliation{Departamento de F\'{\i}sica and Instituto de Computaci\'on Cient\'{\i}fica Avanzada (ICCAEx), Universidad de Extremadura, E-06006 Badajoz, Spain}

\begin{abstract}

The Navier--Stokes transport coefficients of multicomponent granular suspensions at moderate densities are obtained in the context of the (inelastic) Enskog kinetic theory. The suspension is modeled as an ensemble of solid particles where the influence of the interstitial gas on grains is via a viscous drag force plus a stochastic Langevin-like term defined in terms of a background temperature. In the absence of spatial gradients, it is shown first that the system reaches a homogeneous steady state where the energy lost by inelastic collisions and viscous friction is compensated for by the energy injected by the stochastic force. Once the homogeneous steady state is characterized, a \emph{normal} solution to the set of Enskog equations is obtained by means of the Chapman--Enskog expansion around the \emph{local} version of the homogeneous state. To first-order in spatial gradients, the Chapman--Enskog solution allows us to identify the Navier--Stokes transport coefficients associated with the mass, momentum, and heat fluxes. In addition, the first-order contributions to the partial temperatures and the cooling rate are also calculated. Explicit forms for the diffusion coefficients, the shear and bulk viscosities, and the first-order contributions to the partial temperatures and the cooling rate are obtained in steady-state conditions by retaining the leading terms in a Sonine polynomial expansion. The results show that the dependence of the transport coefficients on inelasticity is clearly different from that found in its granular counterpart (no gas phase). The present work extends previous theoretical results for \emph{dilute} multicomponent granular suspensions [Khalil and Garz\'o, Phys. Rev. E \textbf{88}, 052201 (2013)] to higher densities.

\end{abstract}

\draft
\date{\today}
\maketitle

\section{Introduction}
\label{sec1}

It is known that granular matter in nature is generally immersed in a fluid, like air or water, and so a granular flow is a multiphase process. However, there are situations where the influence of the interstitial fluid on the granular flow can be ignored. This happens for instance when the stress due to the grains is greater than that exerted by the fluid phase. Otherwise, there are many interesting phenomena (such as species segregation in granular mixtures \cite{MLNJ01,NSK03,SSK04,WXZS08,CPSK10,PGM14}) where the effect of the fluid phase cannot be neglected and hence, in principle, one has to start from a theoretical description that accounts for both phases (fluid and solid phases). In the case of monodisperse gas-solid flows, one possibility would be to describe the granular suspension in terms of a set of two coupled kinetic equations for each one of the velocity distributions of the different phases. However, the resulting theory would be very difficult to solve, since in particular the different phases evolve over quite different spatial and temporal scales. The problem would be even more complex when one considers multicomponent gas-solid flows. Thus, due to the technical difficulties involved in the above approach, it is more frequent in gas-solid flows to consider a suspension model where the effect of the interstitial fluid on the solid particles is via an effective external force \cite{KH01}.

The fluid-solid external force that models the effect of the viscous gas on solid particles is usually constituted by two different terms \cite{GSVP11bis,GTSH12,KG13,HTG17}. On the one hand, the first term includes a dissipative force obeying Stokes' law, namely a viscous drag force proportional to the instantaneous particle velocity. On the other hand, the second term has a stochastic component, modeled as a Gaussian white noise \cite{K81}. This stochastic force provides kinetic energy to the solid particles by randomly kicking them \cite{WM96}. Hence, while the drag force tries to model the friction of grains with the interstitial gas phase, the stochastic Langevin-like term mimics the energy transfer from the surrounding gas particles to the granular particles. The above suspension model has been recently \cite{GGG19a} employed to get the Navier--Stokes transport coefficients of monocomponent granular suspensions by solving the Enskog equation for inelastic hard spheres by means of the Chapman--Enskog method \cite{CC70} adapted to dissipative dynamics.

The determination of the Navier--Stokes transport coefficients of multicomponent granular suspensions is challenging. This target is not only relevant from a fundamental point of view but also from a more practical point of view since real gas-solid flows are usually present in nature as an ensemble of particles of different masses, sizes, and coefficients of restitution. In such case, given that the number of variables and parameters involved in the analysis of multicomponent systems is very large, it is usual to consider first more simple systems, such as multicomponent granular suspensions at low density. This was carried out previously in three different papers \cite{KG13,KG18,KG19} where the complete set of Navier--Stokes transport coefficients of a binary mixture were obtained from the Boltzmann kinetic equation.

The objective of this paper is to extend the analysis performed for \emph{dilute} bidisperse suspensions \cite{KG13,KG18,KG19} to the (inelastic) Enskog kinetic theory \cite{G19} for a description of hydrodynamics and transport at higher densities. Since this theory applies for moderate densities (let's say for instance, solid volume fraction $\phi\lesssim 0.25$ for hard spheres), a comparison between kinetic theory and molecular dynamics (MD) simulations becomes practical. This is perhaps one of the main motivations of the present study.

As mentioned before, we want to derive here the Navier--Stokes hydrodynamic equations of multicomponent granular suspensions by solving the Enskog kinetic equation by the Chapman--Enskog method. An important point in the application of this method to the Enskog equation is the reference state to be used in the perturbation scheme. As in the case of \emph{dry} (no gas phase) granular gases \cite{G19}, the zeroth-order velocity distribution $f_i^{(0)}$ of the component $i$ cannot be chosen a \emph{priori} and must be consistently obtained as a solution of the Enskog equation in the absence of spatial gradients. Since we are interested here in computing the transport coefficients under steady state conditions, for simplicity one could  take a steady distribution $f_i^{(0)}$ at any point of the system \cite{GM02,G11a}. However, this steady distribution is not the most general election for $f_i^{(0)}$ since the presence of the interstitial fluid introduces the possibility of a local energy unbalance and hence, the zeroth-order distributions $f_i^{(0)}$ of each component in the Chapman--Enskog solution are not in general stationary distributions. This is because for arbitrary small deviations from the homogeneous steady state the energy gained by grains due to collisions with the background fluid cannot be locally compensated with the other cooling terms, arising from the viscous friction and the collisional dissipation. Thus, in order to get the transport coefficients, we have to achieve first the \emph{unsteady} set of integral equations verifying the first-order distributions $f_i^{(1)}$ and then, we have to solve the above set under steady-state conditions. As a consequence, the transport coefficients depend not only on the steady temperature but also on some quantities (derivatives of the temperature ratio) which provide an indirect information on the departure of the time-dependent solution $f_i^{(0)}$ from its stationary form.

The mass, momentum, and heat fluxes are calculated here up to first order in the spatial gradients of the hydrodynamic fields. In addition, there are contributions to the partial temperatures and the cooling rate proportional to the divergence of the flow velocity field. These new contributions have been recently \cite{GGG19b} evaluated for dry granular mixtures. As happens for binary systems \cite{GDH07,GHD07,MGH12}, the determination of the twelve relevant Navier--Stokes transport coefficients of a binary mixture (ten transport coefficients and two first-order contributions to the partial temperatures and the cooling rate) requires to solve ten coupled integral equations. This is of course a very long task. For this reason, in this work we will address the determination of the four diffusion coefficients associated with the mass flux, the shear and bulk viscosities coefficients and the first-order contributions to the partial temperatures and the cooling rate.

The plan of the paper is as follows. The set of coupled Enskog equations for multicomponent granular suspensions and the corresponding balance equations for the densities of mass, momentum, and energy are derived in Sec.\ \ref{sec2}. Then, Sec.\ \ref{sec3} analyzes the steady homogeneous state. As in previous works \cite{KG13,GMT12,ChVG13}, scaling solutions are proposed whose dependence on the temperature $T$ occurs through the dimensionless velocity $\mathbf{c}=\mathbf{v}/v_0$ ($v_0$ being a thermal speed) and the reduced temperature $\theta=T/T_\text{ex}$ ($T_\text{ex}$ being the background temperature). Theoretical predictions for the temperature ratio of both components $T_1/T_2$ are compared against MD simulations. The comparison shows in general a good agreement for conditions of practical interest. Section \ref{sec4} is focused on the application of the Chapman--Enskog expansion around the unsteady reference distributions $f_i^{(0)}(\mathbf{r},\mathbf{v},t)$ up to first order in the spatial gradients. The Navier--Stokes transport coefficients are defined in Sec.\ \ref{sec5} and given in terms of the solutions of a set of linear coupled integral equations. The leading terms in a Sonine polynomial expansion are considered in Sec.\ \ref{sec6} to solve the integral equations defining the diffusion coefficients, the shear viscosity, and the first-order contributions to the partial temperatures and the cooling rate. All these coefficients are explicitly determined as functions of both the granular and background temperatures, the density, the concentration, and the mechanical parameters of the mixture (masses, diameters, and coefficients of restitution). The dependence of the transport coefficients, the partial temperatures, and the cooling rate on the parameter space is illustrated in Sec.\ \ref{sec7} for several systems. It is shown that the impact of the gas phase on the transport coefficients is in general quite significant since their dependence on inelasticity is different from the one obtained for dry granular mixtures \cite{G19,GDH07,GHD07,MGH12}. The paper is closed in Sec.\ \ref{sec8} with a brief discussion of the results obtained in this work. Further details of the calculations carried out here are given in three Appendices.

\section{Enskog kinetic equation for polidisperse gas-solid flows}
\label{sec2}

\subsection{Model for multicomponent granular suspensions}

We consider a binary mixture composed by inelastic hard disks ($d=2$) or spheres ($d=3$) of masses $m_i$ and diameters $\sigma_i$ ($i=1,2$). The solid particles are immersed in an ordinary gas of viscosity $\eta_g$. Spheres are assumed to be completely smooth so that, inelasticity of collisions between particles of the component $i$ with particles of the component $j$ is only characterized by the constant (positive) coefficients of restitution $\al_{ij}\leqslant 1$. The mixture is also assumed to be in the presence of the gravitational field and hence, each particle feels the action of the force $\mathbf{F}_i=m_i \mathbf{g}$, where $\mathbf{g}$ is the gravity acceleration. For moderate densities, the one-particle velocity distribution function $f_i(\mathbf{r}, \mathbf{v}, t)$ of the component $i$ verifies the set of nonlinear Enskog equations \cite{G19}
\beq
\label{2.1}
\frac{\partial f_i}{\partial t}+\mathbf{v}\cdot \nabla f_i+\mathbf{g}\cdot \frac{\partial f_i}{\partial \mathbf{v}}+\mathcal{F}_i f_i=\sum_{j=1}^2J_{ij}[\mathbf{r},\mathbf{v}|f_i,f_j],
\eeq
where the Enskog collision operator is
\begin{widetext}
\beqa
\label{2.2}
J_{ij}\left[\mathbf{r}_1, \mathbf{v}_1|f_i,f_j\right]&=&\sigma_{ij}^{d-1}\int \dd\mathbf{v}_2\int \dd\widehat{\boldsymbol{\sigma}}\Theta\left(\widehat{\boldsymbol{\sigma}}\cdot\mathbf{g}_{12}\right)
\left(\widehat{\boldsymbol{\sigma}}\cdot\mathbf{g}_{12}\right)\Big[\alpha_{ij}^{-2}\chi_{ij}
(\mathbf{r}_1,\mathbf{r}_1-\boldsymbol{\sigma}_{ij})f_i(\mathbf{r}_1,\mathbf{v}_1'',t)
f_j(\mathbf{r}_1-\boldsymbol{\sigma}_{ij},\mathbf{v}_2'',t)
\nonumber\\
& &
-\chi_{ij}(\mathbf{r}_1,\mathbf{r}_1+\boldsymbol{\sigma}_{ij})f_i(\mathbf{r}_1,\mathbf{v}_1,t)f_j(\mathbf{r}_1
+\boldsymbol{\sigma}_{ij},\mathbf{v}_2,t)\Big].
\eeqa
\end{widetext}
In Eq.\ \eqref{2.1}, the operator $\mathcal{F}_i$ represents the fluid-solid interaction force that models the effect of the viscous gas on the solid particles of the component $i$. Its explicit form will be given below. In addition, $\boldsymbol{\sigma}_{ij}=\sigma_{ij} \widehat{\boldsymbol{\sigma}}$, $\sigma_{ij}=(\sigma_i+\sigma_j)/2$, $\widehat{\boldsymbol{\sigma}}$ is a unit vector directed along the line of centers from the sphere of the component $i$ to that of the component $j$ at contact, $\Theta$ is the Heaviside step function, $\mathbf{g}_{12}=\mathbf{v}_1-\mathbf{v}_2$ is the relative velocity of the colliding pair, and $\chi_{ij}(\mathbf{r}_1,\mathbf{r}_1+\boldsymbol{\sigma}_{ij})$ is the equilibrium pair correlation function of two hard spheres, one for the component $i$ and the other for the component $j$ at contact (namely, when the distance between their centers is $\sigma_{ij}$). The precollisional velocities $(\mathbf{v}_1'',\mathbf{v}_2'')$ are given by
\begin{equation}
\label{2.3}
\mathbf{v}_1''=\mathbf{v}_1-\mu_{ji}\left(1+\alpha_{ij}^{-1}\right)\left(\boldsymbol{\widehat{\sigma}}
\cdot\mathbf{g}_{12}\right)\boldsymbol{\widehat{\sigma}},
\eeq
\beq
\label{2.3.1}
\mathbf{v}_2'=\mathbf{v}_2+\mu_{ij}\left(1+\alpha_{ij}^{-1}\right)\left(\boldsymbol{\widehat{\sigma}}
\cdot\mathbf{g}_{12}\right)\boldsymbol{\widehat{\sigma}},
\end{equation}
where $\mu_{ij}=m_i/(m_i+m_j)$.

As in previous works on granular suspensions \cite{GTSH12,HTG17,GGG19,GGG19a}, the influence of the surrounding gas on the dynamics of grains is accounted for via an instantaneous fluid force. For low Reynolds numbers, we assume that the external force is composed by two independent terms. One term is a viscous drag force ($\mathbf{F}_i^\text{drag}$) proportional to the particle velocity $\mathbf{v}$. This term takes into account the friction of particles of the component $i$ with the viscous gas. A subtle point in the choice of the explicit form of the drag force $\mathbf{F}_i^\text{drag}$ for multicomponent systems is that it can be defined to be the same for both components or it can be chosen to be different for both components. Here, in consistency with simulations of bidisperse gas-solid flows \cite{YS09a,YS09b,HYS10}, we will assume that
\beq
\label{2.4}
\mathbf{F}_i^\text{drag}=-m_i \gamma_i \left(\mathbf{v}-\mathbf{U}_g\right),
\eeq
where $\gamma_i$ is the (positive) drift or friction coefficient associated with the component $i$. In addition, since the model tries to model gas-solid flows, the drag force \eqref{2.4} has been defined in terms of the relative velocity $\mathbf{v}-\mathbf{U}_g$ where $\mathbf{U}_g$ is the mean fluid velocity of the gas phase. This latter quantity is assumed to be a known quantity of the suspension model. Thus, according to Eq.\ \eqref{2.4}, in the Enskog equation \eqref{2.1} the drag force is represented by the operator
\beq
\label{2.5}
\mathcal{F}_i^\text{drag}f_i\to -\gamma_i \frac{\partial}{\partial \mathbf{v}}\cdot \left(\mathbf{v}-\mathbf{U}_g\right)f_i.
\eeq

The second term in the gas-to-solid force corresponds to a stochastic Langevin force ($\mathbf{F}_i^\text{st}$) representing Gaussian white noise \cite{K81}. This force attempts to simulate the kinetic energy gain of grains due to eventual collisions with the more energetic particles of the surrounding gas \cite{WM96}. In the context of the Enskog equation \eqref{2.1}, the stochastic force is represented by a Fokker--Planck collision operator of the form \cite{NE98,HBB00,DHGD02,BT02}
\beq
\label{2.6}
\mathcal{F}_i^\text{st}f_i\to -\frac{\gamma_i T_\text{ex}}{m_i}\frac{\partial^2 f_i}{\partial v^2},
\eeq
where $T_\text{ex}$ can be interpreted as the temperature of the background (or bath) gas.

Although the drift coefficient $\gamma_i$ is in general a tensor, here for simplicity we assume that this coefficient is a scalar proportional to  the viscosity of the gas phase $\eta_g$ \cite{KH01}. In addition, according to the results obtained in lattice-Boltzmann simulations of low-Reynolds-number fluid flow in bidisperse suspensions \cite{YS09a,YS09b,HYS10}, the friction coefficients $\gamma_i$ must be functions of the partial volume fractions $\phi_i$ and the total volume fraction $\phi=\phi_1+\phi_2$ where
\beq
\label{2.5.1}
\phi_i=\frac{\pi^{d/2}}{2^{d-1}d\Gamma \left(\frac{d}{2}\right)} n_i\sigma_i^d.
\eeq
Here,
\beq
\label{2.10}
n_i=\int \dd \mathbf{v}\; f_i(\mathbf{v})
\eeq
is the local number density of the component $i$. The coefficients $\gamma_i$ can be written as
\beq
\label{2.5.2}
\gamma_i=\gamma_0 R_i(\phi_i,\phi),
\eeq
where $\gamma_0 \propto \eta_g \propto \sqrt{T_\text{ex}}$. Explicit forms of $R_i(\phi_i,\phi)$ can be found in the literature for polydisperse gas-solid flows \cite{YS09a,YS09b,HYS10}. In particular, for hard spheres ($d=3$), low-Reynolds-number fluid and moderate densities, Yin and Sundaresan \cite{YS09b} have proposed the expression $\gamma_i=(18 \eta_g/\rho \sigma_{12}^2)R_i$ where the dimensionless function $R_i$ is
\beqa
\label{2.5.3}
R_i(\phi_i,\phi)&=&\frac{\rho \sigma_{12}^2}{\rho_i \sigma_i^2}\frac{(1-\phi)\phi_i\sigma_i}{\phi}\sum_{j=1}^{2}\frac{\phi_j}{\sigma_j}\Bigg[
\frac{10\phi}{\left(1-\phi\right)^2}
\nonumber\\
& & +\left(1-\phi\right)^2\left(1+1.5\sqrt{\phi}\right)\Bigg].
\eeqa

Hence, according to Eqs.\ \eqref{2.5} and \eqref{2.6}, the operator $\mathcal{F}_i f_i$ reads
\beqa
\label{2.7}
\mathcal{F}_i f_i&=& \mathcal{F}_i^\text{drag}f_i+\mathcal{F}_i^\text{st}f_i\nonumber\\
& \to & -\gamma_i \Delta \mathbf{U}\cdot\frac{\partial f_i}{\partial\mathbf{v}}-\gamma_i\frac{\partial}{\partial\mathbf{v}}\cdot\mathbf{V}f_i-\frac{\gamma_i T_{\text{ex}}}{m_i}\frac{\partial^2 f_i}{\partial v^2},
\nonumber\\
\eeqa
and the Enskog equation for the component $i$ becomes
\beqa
\label{2.8}
& & \frac{\partial f_i}{\partial t}+\mathbf{v}\cdot \nabla f_i+\mathbf{g}\cdot \frac{\partial f_i}{\partial \mathbf{v}}-\gamma_i \Delta \mathbf{U}\cdot \frac{\partial f_i}{\partial\mathbf{v}}-\gamma_i\frac{\partial}{\partial\mathbf{v}}\cdot\mathbf{V}f_i\nonumber\\
& & -\frac{\gamma_i T_{\text{ex}}}{m_i}\frac{\partial^2 f_i}{\partial v^2}=\sum_{j=1}^2\; J_{ij}[f_i,f_j].
\eeqa
In Eq.\ \eqref{2.7}, $\Delta \mathbf{U}=\mathbf{U}-\mathbf{U}_g$, $\mathbf{V}=\mathbf{v}-\mathbf{U}$ is the peculiar velocity, and
\beq
\label{2.9}
\mathbf{U}=\rho^{-1}\sum_{i=1}^2 \int \dd \mathbf{v}\; m_i \mathbf{v} f_i(\mathbf{v})
\eeq
is the local mean flow velocity of the mixture. Here, $\rho=\sum_i \rho_i$ is the total mass density and $\rho_i=m_in_i$ is the mass density of the component $i$.

The suspension model \eqref{2.8} is similar to the one proposed in Ref.\ \cite{KG13} to obtain the Navier--Stokes transport coefficients of multicomponent granular suspensions at low-density. In this latter model \cite{KG13}, the gas phase depends on two parameters, namely the friction coefficient of the drag force ($\gamma_\text{b}$ in the notation of Ref.\ \cite{KG13}) and the strength of the correlation ($\xi_\text{b}^2$ in the notation of Ref.\ \cite{KG13}). However, in contrast with the suspension model proposed here, both parameters ($\gamma_\text{b}$ and $\xi_\text{b}^2$) were assumed to be independent and the same for each one of the components. Therefore, in the low-density limit, the results derived here reduce to those obtained before \cite{KG13} when one makes the changes $\gamma_1=\gamma_2=\gamma_\text{b}$ [with $R_i(\phi_i,\phi)=1$] and $\xi_\text{b}^2= 2 \gamma_\text{b} T_\text{ex}$. Here, in the notation of Ref.\ \cite{KG13}, the other constants of the driven model are chosen to be $\beta=0$ and $\lambda=1$; this is one of the possible elections consistent with the fluctuation-dissipation theorem for elastic collisions \cite{K81}.

\subsection{Balance equations}

Apart from the partial densities $n_i$ and the flow velocity $\mathbf{U}$, the other important hydrodynamic field is the granular temperature $T$. It is defined as
\beq
\label{2.11}
T=\frac{1}{n}\sum_{i=1}^2\int \dd\mathbf{v}\frac{m_{i}}{d}V^{2}f_{i}(\mathbf{v}),
\eeq
where $n=n_{1}+n_{2}$ is the total number density. The granular temperature $T$ can also be defined in terms of the partial kinetic temperatures $T_1$ and $T_2$ of the components $1$ and $2$, respectively. The partial kinetic temperature $T_i$ measures the mean kinetic energy of the component $i$. They are defined as
\begin{equation}
\label{2.12a}
T_i=\frac{m_{i}}{d n_i}\int\; \dd\mathbf{v}\;V^{2}f_{i}(\mathbf{ v}), \quad i=1,2.
\end{equation}
In accordance with Eq.\ \eqref{2.11}, the granular temperature $T$ of the mixture can be also written as
\beq
\label{2.12}
T=\sum_{i=1}^2\, x_i T_i,
\eeq
where $x_i=n_i/n$ is the concentration or mole fraction of the component $i$.

In order to obtain the balance equations for the hydrodynamic fields, an important property of the integrals involving the (inelastic) Enskog collision operator $J_{ij}[\mathbf{r},\mathbf{v}|f_i,f_j]$ is \cite{BP04,G19}
\beqa
\label{2.13}
I_\psi&\equiv & \int\; \dd \mathbf{v}_1\; \psi(\mathbf{v}_1) J_{ij}[\mathbf{r}_1,\mathbf{v}_1|f_i,f_j]\nonumber\\
&=&\sigma_{ij}^{d-1}\int \dd \mathbf{v}_1\int\ \dd{\bf v}_{2}\int \dd\widehat{\boldsymbol{\sigma}}\,
\Theta (\widehat{{\boldsymbol {\sigma }}}\cdot \mathbf{g}_{12})(\widehat{\boldsymbol {\sigma }}\cdot \mathbf{g}_{12})\nonumber\\
& & \times
\chi_{ij}(\mathbf{r}_1,\mathbf{r}_1+\boldsymbol{\sigma}_{ij}) f_i(\mathbf{r}_1,\mathbf{v}_1,t)
f_j(\mathbf{r}_1+\boldsymbol{\sigma}_{ij},\mathbf{v}_2,t)
\nonumber\\
& &\times
\left[\psi(\mathbf{v}_1')-\psi(\mathbf{v}_1)\right],
\eeqa
where $\psi(\mathbf{v})$ is an arbitrary function of $\mathbf{v}$ and
\beq
\label{2.14}
\mathbf{v}_1'=\mathbf{v}_1-\mu_{ji}\left(1+\alpha_{ij}\right)\left(\boldsymbol{\widehat{\sigma}}
\cdot\mathbf{g}_{12}\right)\boldsymbol{\widehat{\sigma}}.
\eeq
The balance equations for the densities of mass, momentum, and energy can be derived by taking velocity moments in the Enskog equation \eqref{2.8} and using the property \eqref{2.13}. They read
\begin{equation}
\label{2.15}
D_t n_i+n_i\nabla\cdot \mathbf{U}+\frac{\nabla\cdot\mathbf{j}_i}{m_i}=0,
\end{equation}
\beq
\label{2.16}
D_t\mathbf{U}+\rho^{-1}\nabla\cdot\mathsf{P}=\mathbf{g}-\rho^{-1}\Delta\mathbf{U}\sum_{i=1}^2\rho_i \gamma_i
-\rho^{-1}\left(\gamma_1-\gamma_2\right)\mathbf{j}_1,
\eeq
\beqa
\label{2.17}
D_tT&-&\frac{T}{n}\sum_{i=1}^2\frac{\nabla\cdot\mathbf{j}_i}{m_i}+\frac{2}{dn}
\left(\nabla\cdot\mathbf{q}+\mathsf{P}:\nabla\mathbf{U}\right)\nonumber\\
& =&-\frac{2}{d n}\Delta\mathbf{U}\cdot \sum_{i=1}^2 \gamma_i \; \mathbf{j}_i+
2\sum_{i=1}^2 x_i \gamma_i\left(T_{\text{ex}}-T_i\right)\nonumber\\
& & -\zeta T.
\eeqa
In the above equations, $D_t=\partial_t+\mathbf{U}\cdot\nabla$ is the  material derivative, and
\beq
\label{2.18}
\mathbf{j}_i=m_i\int\;\dd\mathbf{v}\; \mathbf{V}f_i(\mathbf{v})
\eeq
is the mass flux for the component $i$ relative to the local flow $\mathbf{U}$. A consequence of the definition \eqref{2.18} of the fluxes $\mathbf{j}_i$ is that $\mathbf{j}_1=-\mathbf{j}_2$. The pressure tensor $\mathsf{P}(\mathbf{r},t)$ and the heat flux $\mathbf{q}(\mathbf{r},t)$ have both kinetic and collisional transfer contributions, i.e.,
\beq
\label{2.19}
\mathsf{P}=\mathsf{P}^\text{k}+\mathsf{P}^\text{c}, \quad \mathbf{q}=\mathbf{q}^\text{k}+\mathbf{q}^\text{c}.
\eeq
The kinetic contributions $\mathsf{P}^\text{k}$ and $\mathbf{q}^\text{k}$ are given by
\beq
\label{2.20}
\mathsf{P}^\text{k}=\sum_{i=1}^2\int \dd\mathbf{v}\; m_i\mathbf{V}\mathbf{V}f_i(\mathbf{v}),
\eeq
\beq
\label{2.21}
\mathbf{q}^\text{k}=\sum_{i=1}^2\int \dd\mathbf{v}\; \frac{m_i}{2}V^2\mathbf{V}f_i(\mathbf{v}).
\eeq
The collisional transfer contributions are \cite{GDH07}
\beqa
\label{2.22}
\mathsf{P}^\text{c}&=&\sum_{i=1}^2\sum_{j=1}^2\sigma_{ij}^d m_{ij}\frac{1+\alpha_{ij}}{2}\int \dd\mathbf{v}_1\int \dd\mathbf{v}_2
\int \dd\widehat{\boldsymbol{\sigma}}\nonumber\\
& & \times \Theta\left(\widehat{\boldsymbol{\sigma}}\cdot\mathbf{g}_{12}\right)\left(\widehat{\boldsymbol{\sigma}}
\cdot\mathbf{g}_{12}\right)^2\widehat{\boldsymbol{\sigma}}\widehat{\boldsymbol{\sigma}}\int_{0}^{1}\dd x\nonumber\\
& & \times f^{(2)}_{ij}[\mathbf{r}-x\boldsymbol{\sigma}_{ij},\mathbf{r}+(1-x)\boldsymbol{\sigma}_{ij},\mathbf{v}_1,\mathbf{v}_2,t],
\eeqa
\beqa
\label{2.23}
\mathbf{q}^\text{c}&=&\sum_{i=1}^2\sum_{j=1}^2\sigma_{ij}^dm_{ij}\frac{1+\alpha_{ij}}{8}\int \dd\mathbf{v}_1\int \dd\mathbf{v}_2\int \dd\widehat{\boldsymbol{\sigma}}\nonumber\\
& & \times \Theta\left(\widehat{\boldsymbol{\sigma}}\cdot\mathbf{g}_{12}\right)\left(\widehat{\boldsymbol{\sigma}}
\cdot\mathbf{g}_{12}\right)^2\widehat{\boldsymbol{\sigma}}\Big[4\left(\widehat{\boldsymbol{\sigma}}\cdot\mathbf{G}_{ij}\right)\nonumber\\
& & +
\left(\mu_{ji}-\mu_{ij}\right)\left(1-\alpha_{ij}\right)\left(\boldsymbol{\hat{\sigma}}\cdot\mathbf{g}_{12}\right)\Big]\int_{0}^{1}\dd x\nonumber\\
& &
\times f^{(2)}_{ij}[\mathbf{r}-x\boldsymbol{\sigma}_{ij},\mathbf{r}+(1-x)\boldsymbol{\sigma}_{ij},\mathbf{v}_1,\mathbf{v}_2;t].
\eeqa
Here, $m_{ij}=m_im_j/(m_i+m_j)$ is the reduced mass, $\mathbf{G}_{ij}=\mu_{ij}\mathbf{V}_1+\mu_{ji}\mathbf{V}_2$ is the velocity of center of mass and
\beq
\label{2.24}
f^{(2)}_{ij}(\mathbf{r}_1,\mathbf{r}_2,\mathbf{v}_1,\mathbf{v}_2,t)\equiv\chi_{ij}(\mathbf{r}_1,\mathbf{r}_2)
f_i(\mathbf{r}_1,\mathbf{v}_1,t)f_j(\mathbf{r}_2,\mathbf{v}_2,t).
\eeq
Finally, the (total) cooling rate $\zeta$ is due to inelastic collisions among all components. It is given by
\beqa
\label{2.25}
\zeta&=&\frac{1}{2dnT}\sum_{i=1}^2\sum_{j=1}^2\sigma_{ij}^{d-1}m_{ij}\left(1-\alpha_{ij}^2\right)\int \dd\mathbf{v}_1\int \dd\mathbf{v}_2\int \dd\widehat{\boldsymbol{\sigma}}\nonumber\\
& & \times \Theta\left(\widehat{\boldsymbol{\sigma}}\cdot\mathbf{g}_{12}\right)\left(\widehat{\boldsymbol{\sigma}}
\cdot\mathbf{g}_{12}\right)^3f^{(2)}_{ij}
[\mathbf{r},\mathbf{r}+\boldsymbol{\sigma}_{ij},\mathbf{v}_1,\mathbf{v}_2;t].
\eeqa

As expected, the balance equations \eqref{2.15}--\eqref{2.17} are not a closed set of equations for the fields $n_1$, $n_2$, $\mathbf{U}$, and $T$. To become these equations into a set of closed equations, one has to express the fluxes and the cooling rate in terms of the hydrodynamic fields and their gradients. The corresponding constitutive equations can be obtained by solving the Enskog kinetic equation \eqref{2.8} from the Chapman--Enskog method \cite{CC70} adapted to dissipative dynamics. This will be worked out in Sec.\ \ref{sec4}.

\section{Homogeneous steady states}
\label{sec3}

\subsection{Time-dependent state}

Before considering inhomogeneous states, we will study first the homogeneous problem. This state has been widely analyzed in Ref.\ \cite{KG13} for a similar suspension model. In the homogeneous state, the partial densities $n_i(\mathbf{r},t)\equiv n_i$ are constant, the granular temperature $T(\mathbf{r},t)\equiv T(t)$ is spatially uniform, the gas velocity $\mathbf{U}_g$ is assumed to be uniform, and, with an appropriate selection of the reference frame, both mean flow velocities vanish ($\mathbf{U}=\mathbf{U}_g=\mathbf{0}$). Under these conditions and in the absence of gravity ($\mathbf{g}=\mathbf{0}$), Eq.\ \eqref{2.8} reads
\beq
\label{3.1}
\partial_tf_i-\gamma_i\frac{\partial}{\partial\mathbf{v}}\cdot\mathbf{v}f_i-\frac{\gamma_i T_{\text{ex}}}{m_i}\frac{\partial^2 f_i}{\partial v^2}=\sum_{j=1}^2\;J_{ij}[f_i,f_j],
\eeq
where
\beqa
\label{3.2}
J_{ij}[f_i,f_j]&=&\chi_{ij}\sigma_{ij}^{d-1}\int \dd\mathbf{v}_2\int \dd\widehat{\boldsymbol{\sigma}}\Theta\left(\widehat{\boldsymbol{\sigma}}
\cdot\mathbf{g}_{12}\right)\left(\widehat{\boldsymbol{\sigma}}\cdot\mathbf{g}_{12}\right)\nonumber\\
& & \times \Big[\alpha_{ij}^{-2}f_i(\mathbf{v}_1')f_j(\mathbf{v}_2')-f_i(\mathbf{v}_1)f_j(\mathbf{v}_2)\Big]
\eeqa
is the Boltzmann collision operator multiplied by the (constant) pair correlation function $\chi_{ij}$. For homogeneous states, the fluxes vanish and so the only nontrivial balance equation is that of the temperature \eqref{2.17}:
\beq
\label{3.3}
\partial_t T=2\sum_{i=1}^2x_i \gamma_i\left(T_{\text{ex}}-T_i\right)-\zeta T,
\eeq
where, according to Eq.\ \eqref{2.25}, the expression of $\zeta$ for homogeneous states can be written as
\beqa
\label{3.7}
\zeta&=&\frac{\pi^{(d-1)/2}}{2d\Gamma\left(\frac{d+3}{2}\right)}\frac{1}{nT}\sum_{i=1}^2\sum_{j=1}^2
\sigma_{ij}^{d-1}m_{ij}\chi_{ij}(1-\al_{ij}^2)\nonumber\\
& &\times \int\dd \mathbf{v}_1
\int\dd \mathbf{v}_2\; g_{12}^3\; f_i(\mathbf{v}_1)f_j(\mathbf{v}_2).
\eeqa

Analogously, the evolution equation for the partial temperatures $T_i$ can be obtained from Eq.\ \eqref{3.1}  as
\begin{equation}
\label{3.4}
\partial_tT_i=2\gamma_i\left(T_{\text{ex}}-T_i\right)-\zeta_i T_i,
\end{equation}
where we have introduced the partial cooling rates $\zeta_i$ for the partial temperatures $T_i$. They are defined as
\begin{equation}
\label{2.26}
\zeta_i=-
\frac{m_i}{dn_iT_i}\sum_{j=1}^2\int \dd\mathbf{ v}\; V^{2}J_{ij}[f_{i},f_{j}].
\end{equation}
The total cooling rate $\zeta$ can be rewritten in terms of the partial cooling rates $\zeta_i$ when one takes into account the constraint \eqref{2.12} and the evolution equations \eqref{3.3} and \eqref{3.4}. The result is
\beq
\label{2.26.1}
\zeta=\sum_{i=1}^2\; x_i \tau_i \zeta_i,
\eeq
where $\tau_i=T_i/T$ is the temperature ratio of the component $i$.

As usual, for times longer than the mean free time, the system is expected to reach a hydrodynamic regime where the distributions $f_i$ depend on time through the only time-dependent hydrodynamic field of the problem: the granular temperature $T$ \cite{GD99b}. In this regime,
\begin{equation}
\label{3.5}
\partial_tf_i=\frac{\partial f_i}{\partial T}\partial_tT=\Big[2\sum_{i=1}^2x_i\gamma_i\left(\theta^{-1}-\tau_i\right)-\zeta\Big]T
\frac{\partial f_i}{\partial T},
\end{equation}
and the homogeneous Enskog equation \eqref{3.1} becomes
\beqa
\label{3.6}
& & \Big[2\sum_{i=1}^2x_i\gamma_i\left(\theta^{-1}-\tau_i\right)-\zeta\Big]T\frac{\partial f_i}{\partial T}-\gamma_i\frac{\partial}{\partial\mathbf{v}}\cdot\mathbf{v}f_i\nonumber\\
& &-\frac{\gamma_i T_{\text{ex}}}{m_i}\frac{\partial^2 f_i}{\partial v^2}=\sum_{j=1}^2J_{ij}[f_i,f_j],
\eeqa
where $\theta=T/T_{\text{ex}}$.

\subsection{Steady state}

In the \emph{steady} state ($\partial_t T_i=0$), Eq.\ \eqref{3.4} gives the following set of coupled equations for the (asymptotic) partial temperatures $T_{i,s}$:
\beq
\label{3.10}
2\gamma_i\left(T_{\text{ex}}-T_{i,s}\right)-\zeta_{i,s}T_{i,s}=0,
\eeq
where the subscript $s$ means that all the quantities are evaluated in the steady state. To determine these temperatures one has to get the steady-state solution to Eq.\ \eqref{3.6}. By using the relation
\beq
\label{3.10.1}
2\sum_{i=1}^2x_i\gamma_i\left(\theta^{-1}-\tau_i\right)-\zeta=0,
\eeq
Eq.\eqref{3.6} reads
\beq
\label{3.11}
-\gamma_i\frac{\partial}{\partial\mathbf{v}}\cdot\mathbf{v}f_{i,s}-\frac{\gamma_i T_{\text{ex}}}{m_i}\frac{\partial^2 f_{i,s}}{\partial v^2}
=\sum_{j=1}^2J_{ij}[f_{i,s},f_{j,s}].
\eeq
As shown for dilute driven multicomponent granular gases \cite{KG13}, dimensionless analysis requires that $f_{i,s}$ has the scaled form
\beq
\label{3.12}
f_{i,s}(n_i,\mathbf{v}, \gamma, T_{\text{ex}})=n_i v_{0}^{-d}\varphi_{i,s}(\mathbf{c}, x_1,\gamma_{i,s}^*,\theta_s),
\eeq
where the unknown scaled function $\varphi_{i,s}$ depends on the dimensionless parameters
\beq
\label{3.13}
\mathbf{c}=\frac{\mathbf{v}}{v_{0s}}, \quad \gamma_{i,s}^*=\frac{\gamma_i}{n\sigma_{12}^{d-1}v_{0s}}.
\eeq
Here, $v_{0s}=\sqrt{2T_s/\overline{m}}$ is the thermal speed and $\overline{m}=\sum_i m_i/2$. Note that the time-dependent velocity distribution function $f_i(\mathbf{v},t)$ also admits the scaling form \eqref{3.12}.

The (reduced) drift parameters $\gamma_{i,s}^*$ can be easily expressed in terms of the mole fraction, the volume fractions $\phi_i$ and $\phi$, and the (reduced) temperature $\theta_s$ as
\beq
\label{3.15}
\gamma_{i,s}^*=\lambda_i \theta_s^{-1/2}, \quad \lambda_i=\frac{\sqrt{2}\pi^{d/2}}{2^d d \Gamma\left(\frac{d}{2}\right)}\frac{R_i(\phi_i,\phi)}{\sqrt{T_\text{ex}^*}\sum_j (\sigma_{12}/\sigma_j)^d\phi_j},
\eeq
$T_\text{ex}^*\equiv T_\text{ex}/(\overline{m} \sigma_{12}^2 \gamma_0^2)$ being the (dimensionless) background gas temperature. Note that $\lambda_1/R_1=\lambda_2/R_2$. As expected from previous works \cite{GGG19a,GChV13,GMT13,KG13}, the dependence of the scaled distribution $\varphi_{i,s}$ on the temperature is not only through the dimensionless velocity $\mathbf{c}$ but also through the dimensionless parameter $\theta_s$. This scaling differs from the one assumed in free cooling systems \cite{NE98,GD99a} where all the temperature dependence of $\varphi_{i,s}$ is encoded through $\mathbf{c}$.

The scaling given by Eq.\ \eqref{3.12} is equivalent to the one proposed in Ref.\ \cite{KG13} when one makes the mapping $\xi_\text{s}^*\to 2\lambda \theta_\text{s}^{-3/2}$, where $\lambda_1=\lambda_2=\lambda$ and $R_1=R_2=1$. The dimensionless parameter $\xi_\text{s}^*$ is defined by Eq.\ (34) of Ref.\ \cite{KG13}. Thus, in the particular case of $\lambda_i=\lambda$ and $R_i=1$, the results for homogeneous states are consistent with those derived before \cite{KG13} in the dilute regime ($\phi\to 0$).

In reduced form, Eq.\ \eqref{3.11} can be rewritten as
\beq
\label{3.16}
-\gamma_{i,s}^*\frac{\partial}{\partial\mathbf{c}}\cdot\mathbf{c}\varphi_{i,s}
-\frac{\gamma_{i,s}^*}{2M_i\theta_s}\frac{\partial^2 \varphi_{i,s}}{\partial c^2}=\sum_{j=1}^2\; J_{ij}^*[\varphi_{i,s},\varphi_{j,s}],
\eeq
where $M_i=m_i/\overline{m}$ and $J_{ij}^*=\ell J_{ij}/n_i v_{0s}^{1-d}$, $\ell=1/n\sigma_{12}^{d-1}$ being proportional to the mean free path of hard spheres. The knowledge of the distributions $\varphi_i$ allows us to get the partial temperatures and the partial cooling rates. In the case of elastic collisions ($\al_{ij}=1$), $T_{1,s}=T_{2,s}=T_{s}=T_\text{ex}$ and hence, Eq.\ \eqref{3.16} admits the simple solution $\varphi_{i,s}=\pi^{-d/2} M_i^{d/2} e^{-M_i c^2}$. However, the exact form of the above distributions is not known for inelastic collisions and hence, one has to consider approximate forms for $\varphi_{i,s}$. In particular, previous results derived for driven granular mixtures \cite{BT02,GV12,KG14} have shown that the partial temperatures can be well estimated by using Maxwellian distributions at different temperatures for the scaled distributions $\varphi_{i,s}(\mathbf{c})$:
\beq
\label{3.17}
\varphi_{i,s}(\mathbf{c})\to \varphi_{i,\text{M}}(\mathbf{c})=\pi^{-d/2} \beta_i^{d/2} e^{-\beta_i c^2},
\eeq
where $\beta_i=M_i T_s/T_{i,s}$. The (reduced) cooling rate $\zeta_{i,s}^*=\ell \zeta_{i,s}/v_{0s}$ can be determined by taking the approximation \eqref{3.17} in Eq.\ \eqref{2.26}. The result is
\beqa
\label{3.18}
\zeta_{i,s}^*&=&\frac{4\pi^{\left(d-1\right)/2}}{d\Gamma\left(\frac{d}{2}\right)}
\sum_{j=1}^2\;x_{j}\chi_{ij}\mu_{ji}\left(\frac{\sigma_{ij}}{\sigma_{12}}\right)^{d-1}\left(\frac{\beta_i+\beta_j}
{\beta_i\beta_j}\right)^{1/2}\nonumber\\
& & \times \left(1+\alpha_{ij}\right)\left[1-\frac{\mu_{ji}}{2}\left(1+\alpha_{ij}\right)
\frac{\beta_i+\beta_j}{\beta_j}\right].
\eeqa
The (reduced) partial temperatures $\theta_{i,s}=T_{i,s}/T_\text{ex}$ can be obtained from the steady state condition \eqref{3.10} for $i=1,2$. In reduced form, the equation for $\theta_{i,s}$ can be written as
\beq
\label{3.19}
2\lambda_i \theta_s^{-1/2}\left(1-\theta_{i,s}\right)-\zeta_{i,s}^* \theta_{i,s}=0.
\eeq
Note that Eq.\ \eqref{2.12} imposes the constraint $x_1\theta_{1,s}+x_2\theta_{2,s}=\theta_s$. Substitution of Eq.\ \eqref{3.18} into the set of equations \eqref{3.19} allows us to get the partial temperatures in terms of the concentration $x_1$, the solid volume fraction $\phi$, the (reduced) background temperature $T_\text{ex}^*$, and the mechanical parameters of the mixture (mass and diameter ratios and coefficients of restitution). In the low-density limit, Eq.\ \eqref{3.19} is consistent with the one obtained in Ref.\ \cite{KG13} after making the change $2\lambda_i \theta_s^{-1/2}=\xi_s^*$.

\begin{figure}
\includegraphics[width=0.8\columnwidth]{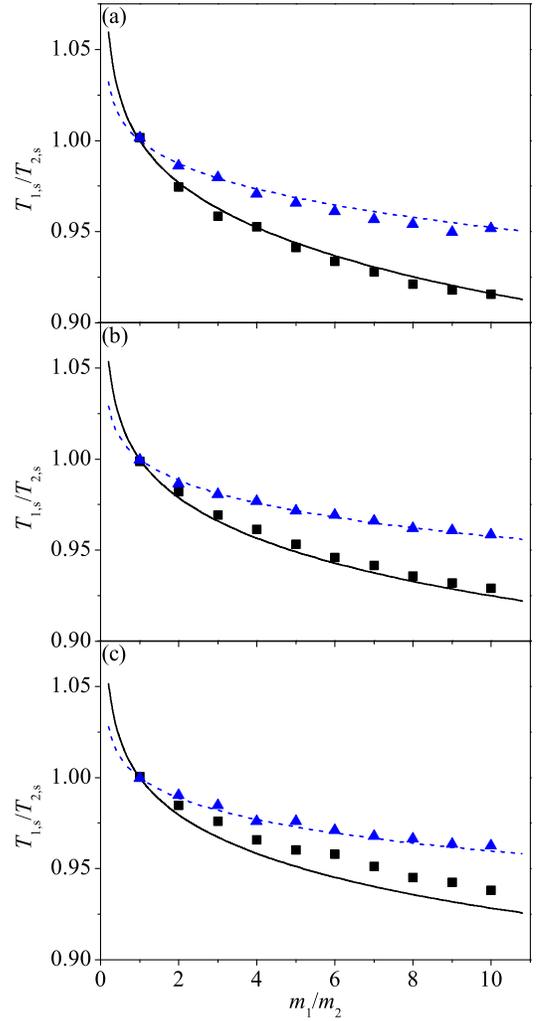}
\caption{Plot of the temperature ratio $T_{1,s}/T_{2,s}$ versus the mass ratio $m_1/m_2$ for a binary mixture of hard spheres ($d=3$) with $x_1=\frac{1}{2}$,  $\sigma_1/\sigma_2=1$, and two different values of the (common) coefficient of restitution $\al$: $\al=0.8$ (solid lines and squares) and $\al=0.9$ (dashed lines and triangles). The lines are the theoretical Enskog results and the symbols refer to the MD simulation results. From top to bottom, the panel (a) corresponds to $\phi=0.00785$, the panel (b) to $\phi=0.1$, and the panel (c) to $\phi=0.2$. The value of the (reduced) background temperature is $T_\text{ex}^*=1$.}
\label{fig1}
\end{figure}

Figure \ref{fig1}  shows the dependence of the temperature ratio $T_{1,s}/T_{2,s}$ on the (common) coefficient of restitution $\al$ ($\al\equiv \al_{11}=\al_{22}=\al_{12})$ for a binary granular suspension of hard spheres ($d=3$). The lines are the theoretical results derived from the Enskog equation while the symbols refer to the results obtained via event-driven MD simulations \cite{L91a,AT05}. We have simulated a system constituted by a total number of $N=20^3$ inelastic, smooth hard spheres. The system is inside a 3D box of size $L$ with periodic boundary conditions. In addition to the inter-particle collisions, particles of each component change their velocities due to the interactions with the bath (with $\mathbf{U}_g=\mathbf{0}$), as explained in Ref.\ \cite{KG14}. Three different values of the solid volume fractions $\phi$ have been analyzed: $\phi=0.00785$, $\phi=0.1$, and $\phi=0.2$. The first system corresponds to a very dilute granular suspension while the two latter can be considered as moderately dense granular suspensions. Two different values of the common coefficient of restitution have been chosen, $\al=0.8$ and $\al=0.9$. Both values of $\al$ represent a moderate degree of inelasticity. The symbols are the simulation data where the squares are for $\al=0.8$ and the triangles are for $\al=0.9$. The Enskog theoretical predictions are given by the solid ($\al=0.8$) and dashed ($\al=0.9$) lines.

Figure \ref{fig1} highlights the excellent agreement found between the Enskog theory and simulations in both the low density limit ($\phi=0.00785$) and moderate density ($\phi=0.1$). This agreement is kept for both values of inelasticity and over the whole range of mass ratios studied. The agreement is also excellent for $\phi=0.2$ and $\al=0.9$; more quantitative discrepancies appear for $\al=0.8$, specially for large values of the mass ratio. These differences between the Enskog theory and MD simulations for moderate densities and strong inelasticity could be due to the fact that the impact of the cumulants (which have been neglected in our solution) on the temperature ratio could be non-negligible in this region of the parameter space or due to a failure of the Enskog theory (namely, molecular chaos hypothesis fails at high densities and strong inelasticity). In any case, the good performance of the Enskog results found here for the temperature ratio contrasts with the results obtained in freely cooling granular mixtures \cite{DHGD02} where significant differences between theory and simulations were found at $\phi=0.2$ (see for instance, Fig.\ 2 of Ref.\ \cite{DHGD02}).

In summary, the comparison performed here for the temperature ratio in homogeneous steady states for granular suspensions shows again that the range of densities for which the Enskog kinetic equation becomes reliable likely decreases with increasing inelasticity. This finding has been already achieved in some previous works \cite{KG14,LBD02,L04,MGAL06,LLC07,MGH14}. However, despite this limitation, the Enskog theory can be still considered as a remarkable theory for describing transport properties for fluids with elastic and inelastic collisions.

\section{Chapman--Enskog solution of the Enskog equation}
\label{sec4}

We assume now that the homogeneous steady state is slightly perturbed by the presence of spatial gradients. These gradients induce fluxes of mass, momentum, and energy. The knowledge of these fluxes allows us to identify the relevant transport coefficients  of the bidisperse granular suspension. As in previous works on granular mixtures \cite{GD02,GDH07,GHD07,KG13}, we consider states that deviate from the reference state (homogeneous time-dependent state) by small spatial gradients. In this situation, the set of Enskog equations \eqref{2.8} can be solved by means of the Chapman--Enskog method \cite{CC70} conveniently adapted to take into account the inelasticity in collisions.

As usual, for times longer than the grain-grain mean free time and distances larger than the grain-grain mean free path, we assume that the granular suspension has reached the so-called hydrodynamic regime \cite{CC70,GS03}. In this regime, (i) the system has completely ``forgotten'' the details of the initial conditions and in addition, (ii) the hydrodynamic description is limited to the bulk domain of the system (namely, a region far away from the boundaries). Under these conditions, the Chapman--Enskog method seeks a special solution to the Enskog kinetic equation: the so-called  \emph{normal} or hydrodynamic solution. This type of solution is characterized by the fact that all space and time dependence of the distributions $f_i(\mathbf{r}, \mathbf{v}, t)$ only occurs via a functional dependence on the hydrodynamic fields.

On the other hand, as noted in previous papers of granular mixtures \cite{GD02,GMD06,GDH07}, there is more flexibility in the choice of the hydrodynamic fields for the mass and heat fluxes of multicomponent granular fluids. Here, to compare with the results previously derived for undriven dense granular mixtures \cite{GDH07}, we take the partial densities $n_1$ and $n_2$, the temperature $T$, and the $d$ components of the local flow velocity $\mathbf{U}$ as the $d+3$ independent fields of the binary mixture. Therefore, in the hydrodynamic regime, the distributions $f_i(\mathbf{r}, \mathbf{v}, t)$ adopt the normal form
\beq
\label{4.1}
f_i(\mathbf{r}, \mathbf{v}, t)=f_i\big[\mathbf{v}|n_1(t), n_2(t), T(t), \mathbf{U}(t)\big].
\eeq
The notation on the right-hand side of Eq.\ \eqref{4.1} indicates a functional dependence on the partial densities, temperature, and flow velocity. Note that the functional dependence means that in order to determine $f_i$ at the point $\mathbf{r}$ we need to know the fields not only at $\mathbf{r}$ but also at the remaining points of the system. This is formally equivalent to knowing $n_1$, $n_2$, $T$, and $\mathbf{U}$ and their spatial derivatives at $\mathbf{r}$.

Since we are perturbing the reference state with small spatial gradients, we can simplify the functional dependence \eqref{4.1} by expanding the distributions $f_i$ in powers of the spatial gradients. In practice, in order to generate this expansion, $f_i$ is expressed as a series expansion in a formal or bookkeeping parameter $\epsilon$:
\beq
\label{4.2}
f_i=f_i^{(0)}+\epsilon f_i^{(1)}+\epsilon^2 f_i^{(2)}+\cdots,
\eeq
where each factor $\epsilon$ means an implicit spatial gradient. Moreover, in ordering the different level of approximations in the Enskog kinetic equation, one has to characterize the magnitude of the friction coefficients $\gamma_i$, the gravity field $\mathbf{g}$, and the term $\Delta \mathbf{U}$ relative to the spatial gradients. As in the case of elastic collisions \cite{CC70}, since the gravity field induces a pressure gradient $\nabla p$ (the so-called barometric formula), it is assumed first that the magnitude of $\mathbf{g}$ is at least of first-order in the perturbation expansion. In addition, since $\gamma_i$ does not give rise to any flux in the mixture, it is considered to be to zeroth-order in gradients. Finally, with respect to the term $\Delta \mathbf{U}$, it is expected that this term is at least to first-order in gradients because $\mathbf{U}$ relaxes to $\mathbf{U}_g$ in the absence of gradients.

As in the conventional Chapman--Enskog method \cite{CC70}, the time derivative $\partial_t$ is also expanded as
\beq
\label{4.3}
\partial_t=\partial_t^{(0)}+\epsilon \partial_t^{(1)}+\cdots.
\eeq
These expansions lead to similar expansions for the Enskog operators $J_{ij}$
\beq
\label{4.3.1}
J_{ij}=J_{ij}^{(0)}+ \epsilon J_{ij}^{(1)}+\cdots,
\eeq
and the fluxes and the cooling rate when substituted into Eqs.\ \eqref{2.18}--\eqref{2.25}:
\beq
\label{4.4}
\mathbf{j}_i=\mathbf{j}_i^{(0)}+\epsilon \mathbf{j}_i^{(1)}+\cdots, \quad \mathsf{P}=\mathsf{P}^{(0)}+\epsilon \mathsf{P}^{(1)}+\cdots,
\eeq
\beq
\label{4.5}
\mathbf{q}=\mathbf{q}^{(0)}+\epsilon \mathbf{q}^{(1)}+\cdots, \quad \zeta=\zeta^{(0)}+\epsilon \zeta^{(1)}+\cdots.
\eeq
In addition, although the partial temperatures $T_i$ are not hydrodynamic quantities, they must be also expanded in powers of the gradients as \cite{KG19,GGG19b}
\beq
\label{4.5.1}
T_i=T_i^{(0)}+\epsilon T_i^{(1)}+\cdots.
\eeq

As usual, the hydrodynamic fields $n_i$, $\mathbf{U}$, and $T$ are defined in terms of the zeroth-order approximation:
\beq
\label{4.5.2}
\int \dd\mathbf{v}\left(f_i-f_i^{(0)}\right)=0,
\eeq
\beq
\label{4.5.2.1}
\sum_{i=1}^2\int \dd\mathbf{v}\; \left\{m_i\mathbf{v}, \frac{m_i}{2}V^2\right\}\left(f_i-f_i^{(0)}\right)=\left\{\mathbf{0},0\right\}.
\eeq
Since the constraints \eqref{4.5.2} and \eqref{4.5.2.1} must hold at any order in $\epsilon$, one has
\beq
\label{4.5.3}
\int \dd\mathbf{v}f_i^{(\ell)}=0,
\eeq
and
\beq
\label{4.5.3.1}
\sum_{i=1}^2\int \dd\mathbf{v}\; \left\{m_i\mathbf{v}, \frac{m_i}{2}V^2\right\}f_i^{(\ell)}=\left\{\mathbf{0},0\right\},
\eeq
for $\ell\geq 1$. A consequence of Eq.\ \eqref{4.5.3.1} is that $\mathbf{j}_1^{(\ell)}=-\mathbf{j}_2^{(\ell)}$ and $n_1 T_1^{(\ell)}=-n_2 T_2^{(\ell)}$ for $\ell\geq 1$. This is the usual application of the Chapman--Enskog method to solve kinetic equations. Here, we will restrict our calculations to first order in $\epsilon$; the so-called Navier--Stokes hydrodynamic order.

\subsection{Zeroth-order approximation}

To zeroth order in $\epsilon$, the Enskog kinetic equation \eqref{2.8} for $f_i^{(0)}$ reads
\beq
\label{4.6}
\partial_t^{(0)}f_i^{(0)}-\gamma_i\frac{\partial}{\partial \mathbf{v}}\cdot \mathbf{V} f_i^{(0)}-\gamma_i \frac{T_{\text{ex}}}{m_i}\frac{\partial^2 f_i^{(0)}}{\partial v^2}=\sum_{j=1}^2\; J_{ij}^{(0)}[f_i^{(0)},f_j^{(0)}],
\eeq
where the collision operator $ J_{ij}^{(0)}[f_i^{(0)},f_j^{(0)}]$ is given by Eq.\ \eqref{3.2} with the replacement $f_i\to f_i^{(0)}(\mathbf{r}, \mathbf{v}, t)$. The balance equations at this order give
\beq
\label{4.6.1}
\partial_t^{(0)}n_i=0, \quad \partial_t^{(0)}\mathbf{U}=\mathbf{0},
\eeq
and
\beq
\label{4.7}
\partial_t^{(0)}T=2\sum_{i=1}^2x_i\gamma_i\left(T_{\text{ex}}-T_i^{(0)}\right)-\zeta^{(0)}T,
\eeq
where $\zeta^{(0)}$ is determined by Eq. \eqref{3.7} to zeroth order. In terms of $\zeta_i^{(0)}$, $\zeta^{(0)}$ is given by Eq.\ \eqref{2.26.1}. An accurate estimate of $\zeta_i^{(0)}$ is obtained by considering the Maxwellian approximation \eqref{3.17} to $\varphi_i$. In this case, $\zeta_i^{(0)}=v_0 \zeta_{i,0}^*/\ell$ where $v_0(T)=\sqrt{2T/\overline{m}}$ and $\zeta_{i,0}^*$ is given by Eq.\ \eqref{3.18} with the replacements $x_i\to x_i(\mathbf{r},t)$, $\chi_{ij}\to \chi_{ij}^{(0)}(\mathbf{r},t)$, $T_{i,s}\to T_i^{(0)}(\mathbf{r},t)$, and $T_s\to T(\mathbf{r},t)$. Here, $\chi_{ij}^{(0)}$ is obtained from the functional $\chi_{ij}\left(\mathbf{r},\mathbf{r}\pm\boldsymbol{\sigma}_{ij}|\left\{n_\ell\right\}\right)$ by evaluating all the densities $n_\ell$ at the point of interest $\mathbf{r}$. Furthermore, in Eqs.\ \eqref{4.6.1} and \eqref{4.7}, use has been made of the isotropy in velocity of the zeroth-order distributions $f_i^{(0)}$ which lead to $\mathbf{j}_i^{(0)}=\mathbf{q}^{(0)}=\mathbf{0}$ and $P_{\lambda\beta}=p\delta_{\lambda\beta}$, where the hydrostatic pressure $p$ is \cite{GDH07}
\beq
\label{4.8}
p=n T+\frac{\pi^{d/2}}{d\Gamma\left(\frac{d}{2}\right)}\sum_{i=1}^2\sum_{j=1}^2\; \mu_{ji}n_in_j \sigma_{ij}^d \chi_{ij}^{(0)} T_i^{(0)}(1+\al_{ij}).
\eeq

Since $f_i^{(0)}$ is a normal solution and the zeroth-order time derivatives of $n_i$ and $\mathbf U$ are zero, then $\partial_t^{(0)}f_i^{(0)}=(\partial_T f_i^{(0)})\partial_t^{(0)}T$ where $\partial_t^{(0)}T$ is given by Eq.\ \eqref{4.7}. With this result, Eq.\ \eqref{4.6} can be rewritten as
\beqa
\label{4.9}
\Lambda^{(0)}T\frac{\partial f_i^{(0)}}{\partial T}&-&\gamma_i\frac{\partial}{\partial\mathbf{v}}\cdot\mathbf{V}f_i^{(0)}-\frac{\gamma_i T_{\text{ex}}}{m_i}\frac{\partial^2 f_i^{(0)}}{\partial v^2}\nonumber\\
& & =\sum_{j=1}^2J_{ij}^{(0)}[f_i^{(0)},f_j^{(0)}],
\eeqa
where
\beq
\label{4.10}
\Lambda^{(0)}\equiv 2\sum_{i=1}^2x_i\gamma_i\left(\theta^{-1}-\tau_i\right)-\zeta^{(0)}.
\eeq
Although Eq.\ \eqref{4.9} has the same form as the one corresponding Enskog equation \eqref{3.6} for a strictly homogeneous state, the zeroth-order solution $f_i^{(0)}(\mathbf{r},\mathbf{v},t)$ is a local distribution function. In fact, the stationary solution to Eq.\ \eqref{4.9} corresponds to $\Lambda^{(0)}=0$ and has been previously studied in Sec.\ \ref{sec3}. However, as noted in previous works \cite{GChV13,KG13,GGG19a,KG19a}, since the densities $n_i(\mathbf{r},t)$ and the granular temperature $T(\mathbf{r},t)$ are defined separately in the local reference state $f_i^{(0)}$, then the temperature is in general a time-dependent function ($\partial_t^{(0)}T \neq 0$). Thus, the distribution $f_i^{(0)}$ depends on time through its dependence on the temperature.

The solution to Eq.\ \eqref{4.9} can be expressed in the form \eqref{3.12} (with the replacements $\gamma_{i,s}^*\to \gamma_i^*$ and $\theta_s \to \theta$) where the scaled distribution $\varphi_i$ verifies the \emph{unsteady} equation
\beqa
\label{4.11}
& & \bigg[2\sum_{i=1}^2x_i\gamma_i^*\left(\theta^{-1}-\tau_i\right)-\zeta_0^*\bigg]
\theta\frac{\partial\varphi_i}{\partial\theta}
+\bigg[\frac{\zeta_0^*}{2}-\sum_{i=1}^2x_i\gamma_i^*\nonumber\\
& & \times \left(\theta^{-1}-\tau_i \right)-\gamma_i^*\bigg]\frac{\partial}{\partial \mathbf{c}}\cdot \mathbf{c}\varphi_i
-\frac{\gamma_i^*}{2M_i\theta}\frac{\partial^2 \varphi_i}{\partial c^2}\nonumber\\
& & =\sum_{j=1}^2J_{ij}^*[\varphi_i,\varphi_j].
\eeqa
Here, the derivative $\partial \varphi_i/\partial \theta$ is taken at constant $\mathbf{c}$, $\zeta_0^*=\ell \zeta^{(0)}/v_0$, $\gamma_i^*=\lambda_i \theta^{-1/2}$, and upon deriving Eq.\ \eqref{4.11} use has been of the property
\beq
\label{4.12}
T\frac{\partial f_i^{(0)}}{\partial T}=-\frac{1}{2}\frac{\partial}{\partial \mathbf{V}}\cdot\mathbf{V}f_i^{(0)}+n_iv_0^{-d}\theta\frac{\partial \varphi_i}{\partial \theta}.
\eeq
The evolution of the temperature ratios $\tau_i$ may be easily obtained by multiplying Eq.\ \eqref{4.11} by $c^2$ and integrating over $\mathbf{c}$. In compact form, the result can be written as
\beq
\label{4.13}
\Lambda^* \theta \frac{\partial \tau_i}{\partial \theta}=-\tau_i \Lambda^*+\Lambda_i^*,
\eeq
where $\tau_i=T_i^{(0)}/T$,
\beq
\label{4.13.1}
\Lambda^*=\frac{\ell \Lambda^{(0)}}{v_0}=x_1\Lambda_1^*+x_2\Lambda_2^*,
\eeq
and
\beq
\label{4.14}
\Lambda_i^*=2\gamma_i^*\left(\theta^{-1}-\tau_i \right)-\tau_i \zeta_{i,0}^*.
\eeq

In the steady state ($\Lambda^*=\Lambda_i^*=0$), Eqs.\ \eqref{4.13} are consistent with Eqs.\ \eqref{3.19} for $i=1,2$. Beyond the steady state, Eq.\ \eqref{4.13} must be numerically solved to obtain the dependence of $\tau_1$ and $\tau_2$ on $\theta$. In addition, as will be shown in Sec.\ \ref{sec5}, to determine the diffusion transport coefficients in the steady state one needs to know the derivatives $\Delta_{\theta,1}\equiv (\partial \tau_1/\partial \theta)_s$, $\Delta_{\lambda_1,1}\equiv (\partial \tau_1/\partial \lambda_1)_s$, $\Delta_{x_1,1}\equiv (\partial \tau_1/\partial x_1)_s$, and $\Delta_{\phi,1}\equiv (\partial \tau_1/\partial \phi)_s$. Here, as before, the subscript $s$ means that all the derivatives are evaluated at the steady state. Since $\lambda_2=(R_2/R_1)\lambda_1$, then $(\partial \tau_1/\partial \lambda_2)_s=(R_1/R_2)(\partial \tau_1/\partial \lambda_1)_s$. Analytical expressions of these derivatives are provided in the Appendix \ref{appA}.

\begin{figure}
\includegraphics[width=0.8\columnwidth]{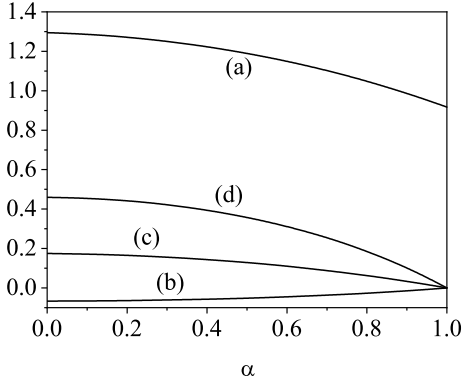}
\caption{Plot of the derivatives $\Delta_{\theta,1}$ (a), $\Delta_{\lambda_1,1}$ (b), $\Delta_{x_1,1}$ (c), and $\Delta_{\phi,1}$ (d) for $d=3$, $x_1=\frac{1}{2}$, $m_1/m_2=10$, $\sigma_1/\sigma_2=1$, $\phi=0.2$, and $T_\text{ex}^*=0.1$.}
\label{fig4}
\end{figure}

The dependence of the derivatives $\Delta_{\theta,1}$, $\Delta_{\lambda_1,1}$, $\Delta_{x_1,1}$, and $\Delta_{\phi,1}$ on the common coefficient of restitution $\al_{ij}\equiv \al$ is plotted in Fig.\ \ref{fig4}. We have considered a three-dimensional system ($d=3$) with $x_1=\frac{1}{2}$, $m_1/m_2=10$, $\sigma_1/\sigma_2=1$, $\phi=0.2$, and $T_\text{ex}^*=0.1$. We observe that in general the magnitude of the derivatives
is not negligible, specially the derivatives $\Delta_{\theta,1}$ and $\Delta_{\phi,1}$ at strong inelasticity.

\subsection{First-order approximation}

The analysis to first order in spatial gradients is more complex than that of the zeroth order. It follows similar steps as those worked out for undriven dense granular mixtures \cite{GDH07,GHD07} and driven dilute granular mixtures \cite{KG13}. Some technical details are displayed in the Appendix \ref{appB} for the sake of completeness. The first-order velocity distribution functions $f_i^{(1)}$ are given by
\beqa
\label{4.15}
f_i^{(1)}&=&\boldsymbol{\mathcal{A}}_i\cdot\nabla\ln T+\sum_{j=1}^2\boldsymbol{\mathcal{B}}_{ij}\cdot \nabla\ln n_j+ \mathcal{C}_{i,\lambda\beta}\frac{1}{2}\Big(\partial_\lambda U_\beta\nonumber\\
& & +\partial_\beta U_\lambda-\frac{2}{d}\delta_{\lambda\beta}\nabla\cdot\mathbf{U}\Big)+\mathcal{D}_i\nabla\cdot\mathbf{U}+\boldsymbol{\mathcal{E}}_i\cdot \Delta\mathbf{U},
\nonumber\\
\eeqa
where $\partial_\beta\equiv \partial/\partial r_\beta$. The unknowns $\boldsymbol{\mathcal{A}}_i(\mathbf{V})$, $\boldsymbol{\mathcal{B}}_{ij}(\mathbf{V})$, $\mathcal{C}_{i,\lambda\beta}(\mathbf{V})$, $\mathcal{D}_i(\mathbf{V})$, and $\boldsymbol{\mathcal{E}}_i(\mathbf{V})$ are functions of the peculiar velocity and they are the solutions of the linear integral equations \eqref{n1}--\eqref{n5}.

On the other hand, as already pointed out in previous works \cite{GChV13,KG13,GGG19a}, the evaluation of the transport coefficients under \emph{unsteady} conditions requires one to know the complete time dependence of the first-order corrections to the mass, momentum, and heat fluxes. This is quite an intricate problem. A more tractable situation occurs when one is interested in evaluating the transport coefficients in steady-state conditions. In this case, since the fluxes $\mathbf{j}_1^{(1)}$, $P_{\lambda\beta}^{(1)}$, and $\mathbf{q}^{(1)}$ are of first order in gradients, then the transport coefficients must be determined to zeroth order in the deviations from the steady state (namely, when the condition $\Lambda^{(0)}=0$ applies). In this situation, the set of  coupled linear integral
equations \eqref{n1}--\eqref{n5} becomes, respectively,
\begin{widetext}
\beqa
\label{4.16}
& &-\left[2\sum_{j=1}^2 \gamma_j x_j\left(\theta^{-1}+\theta\frac{\partial\tau_j}{\partial\theta}\right)
+\frac{1}{2}\zeta^{(0)}+\zeta^{(0)} \theta
\frac{\partial \ln \zeta_0^*}{\partial \theta}\right]\boldsymbol{\mathcal{A}}_i-\gamma_i\frac{\partial}{\partial\mathbf{v}}\cdot\mathbf{V}\boldsymbol{\mathcal{A}}_i
-\frac{\gamma_i T_{\text{ex}}}{m_i}\frac{\partial^2}{\partial v^2}\boldsymbol{\mathcal{A}}_i\nonumber\\
& & +\left(\gamma_2-\gamma_1\right) D_1^T\frac{\partial f_i^{(0)}}{\partial\mathbf{V}}-\sum_{j=1}^2\left(J_{ij}^{(0)}\left[\boldsymbol{\mathcal{A}}_i,f_j^{(0)}\right]
+J_{ij}^{(0)}\left[f_i^{(0)},\boldsymbol{\mathcal{A}}_j\right]\right)=\mathbf{A}_i,
\eeqa
\beqa
\label{4.17}
& & -\gamma_i\frac{\partial}{\partial\mathbf{v}}\cdot\mathbf{V}\boldsymbol{\mathcal{B}}_{ij}
-\frac{\gamma_i T_{\text{ex}}}{m_i}\frac{\partial^2}{\partial v^2}\boldsymbol{\mathcal{B}}_{ij}+\left(\gamma_2-\gamma_1\right)\frac{m_1\rho_j}{\rho^2}D_{1j}\frac{\partial f_i^{(0)}}{\partial\mathbf{V}}-\sum_{\ell=1}^2\left(J_{i\ell}^{(0)}\left[\boldsymbol{\mathcal{B}}_{ij},
f_\ell^{(0)}\right]+J_{i\ell}^{(0)}\left[f_i^{(0)},\boldsymbol{\mathcal{B}}_{\ell j}\right]\right)=\mathbf{B}_{ij}\nonumber\\
& & +\left[n_j\frac{\partial\zeta^{(0)}}{\partial n_j}-2n_j\sum_{\ell=1}^2\left\{\gamma_\ell x_\ell\left[ \left(\theta^{-1}-\tau_\ell\right)\left(\frac{\partial \ln \gamma_\ell}{\partial n_j}+\frac{\partial \ln x_\ell}{\partial n_j}\right)-\left(\frac{\partial\tau_\ell}{\partial x_1}\frac{\partial x_1}{\partial n_j}+\frac{\partial\tau_\ell}{\partial\lambda_1}\frac{\partial\lambda_1}{\partial n_j}+\frac{\partial\tau_\ell}{\partial\phi}\frac{\partial\phi}{\partial n_j}\right)\right]\right\}\right]\boldsymbol{\mathcal{A}}_i,
\eeqa
\beq
\label{4.18}
-\gamma_i\frac{\partial}{\partial\mathbf{v}}\cdot\mathbf{V}\mathcal{C}_{i,\lambda\beta}
-\frac{\gamma_i T_{\text{ex}}}{m_i}\frac{\partial^2}{\partial v^2}\mathcal{C}_{i,\lambda\beta}-\sum_{j=1}^2\left(J_{ij}^{(0)}\left[\mathcal{C}_{i,\lambda\beta},f_j^{(0)}\right]
+J_{ij}^{(0)}\left[f_i^{(0)},\mathcal{C}_{j,\lambda\beta}\right]\right)=C_{i,\lambda\beta},
\eeq
\beq
\label{4.19}
-\gamma_i\frac{\partial}{\partial\mathbf{v}}\cdot\mathbf{V}\mathcal{D}_i-\frac{\gamma_i T_{\text{ex}}}{m_i}\frac{\partial^2}{\partial v^2}\mathcal{D}_i-\left(\zeta^{(1,1)}T+2\sum_{j=1}^2 \gamma_j x_j \varpi_j\right)\frac{\partial f_i^{(0)}}{\partial T}-\sum_{j=1}^2\left(J_{ij}^{(0)}\left[\mathcal{D}_i,f_j^{(0)}\right]+J_{ij}^{(0)}
\left[f_i^{(0)},\mathcal{D}_j\right]\right)=D_i,
\eeq
\beq
\label{4.19.1}
-\gamma_i\frac{\partial}{\partial\mathbf{v}}\cdot\mathbf{V}\boldsymbol{\mathcal{E}}_i-\frac{\gamma_i T_{\text{ex}}}{m_i}\frac{\partial^2}{\partial v^2}\boldsymbol{\mathcal{E}}_i+\rho^{-1}\left(\gamma_2-\gamma_1\right)D^U_1\frac{\partial f_i^{(0)}}{\partial \mathbf{V}}-\sum_{j=1}^2\left(J_{ij}^{(0)}\left[\boldsymbol{\mathcal{E}}_i,f_j^{(0)}\right]+J_{ij}^{(0)}
\left[f_i^{(0)},\boldsymbol{\mathcal{E}}_j\right]\right)=\mathbf{E}_i.
\eeq
The explicit forms of the coefficients $\mathbf{A}_i$, $\mathbf{B}_{ij}$, $C_{i,\lambda\beta}$, $D_i$, and $\mathbf{E}_i$ are given by Eqs.\ \eqref{b11}--\eqref{b14a}, respectively.  These coefficients are functions of $\mathbf{V}$ and the hydrodynamic fields.
\end{widetext}

Upon writing Eqs.\  \eqref{4.16}, \eqref{4.17}, and \eqref{4.19.1}, use has been made of the constitutive equation of the mass flux $\mathbf{j}_1^{(1)}$ to first-order in spatial gradients:
\beq
\label{5.1}
\mathbf{j}_i^{(1)}=-\sum_{j=1}^2\; \frac{m_i\rho_j}{\rho}D_{ij}\nabla\ln n_j-\rho D_i^T\nabla\ln T-D^U_i\Delta\mathbf{U}.
\eeq
In Eq.\ \eqref{5.1}, $D_{ij}$ are the mutual diffusion coefficients, $D_i^T$ are the thermal diffusion coefficients, and $D^U_i$ are the velocity diffusion coefficients. Since $\mathbf{j}_1^{(1)}=-\mathbf{j}_2^{(1)}$, then $D_{21}=-(m_1/m_2) D_{11}$, $D_{22}=-(m_1/m_2) D_{12}$, $D_2^T=-D_1^T$, and $D_2^U=-D_1^U$. In addition, the form of the first-order contribution $\zeta^{(1)}$ to the cooling rate has been also employed to obtain Eq.\ \eqref{4.19}. This coefficient can be written as
\beq
\label{4.20}
\zeta^{(1)}=\zeta_U \nabla \cdot \mathbf{U},
\eeq
where
\beq
\label{4.20.1}
\zeta_U=\zeta^{(1,0)}+\zeta^{(1,1)}.
\eeq
The coefficient $\zeta^{(1,0)}$ is defined by Eq.\ \eqref{b8} while $\zeta^{(1,1)}$ is a functional of the unknowns $\mathcal{D}_i$. Its form is given by Eq.\ \eqref{b16}. Also, in Eq.\ \eqref{4.19}, use has been made of the first-order contribution to the partial temperatures $T_1^{(1)}=-n_2 T_2^{(1)}/n_1$. Since $T_i^{(1)}$ is a scalar, it is coupled to $\nabla \cdot \mathbf{U}$ and has the form \cite{KG19,GGG19b}
\beq
\label{4.20T1}
T_i^{(1)}=\varpi_i \nabla \cdot \mathbf{U},
\eeq
where
\beq
\label{4.20T2}
\varpi_i=\frac{m_i}{d n_i}\int\; \dd \mathbf{v}\;  V^2 \mathcal{D}_i(\mathbf{V}).
\eeq

The direct integration of Eqs.\ \eqref{b11}--\eqref{b14a} for the functions $\mathbf{A}_i$, $\mathbf{B}_{ij}$, $C_{i,\lambda\beta}$, $D_i$, and $\mathbf{E}_i$ yields the following conditions:
\beq
\label{4.21}
\int \dd\mathbf{v} \left(\mathbf{A}_i,\mathbf{B}_{ij},C_{i,\lambda\beta},
D_i,\mathbf{E}_i\right)=\left(\mathbf{0},\mathbf{0},0,0,\mathbf{0}\right),
\eeq
\beq
\label{4.22}
\sum_{i=1}^2\int \dd\mathbf{v} m_i V_\mu
\left(
\begin{array}{ccccc}
A_{i,\lambda}\\
B_{i,\lambda}\\
C_{i,\lambda\beta}\\
D_i\\
E_{i,\lambda}
\end{array}
\right)
=
\left(
\begin{array}{ccccc}
0\\
0\\
0\\
0\\
0
\end{array}
\right),
\eeq
\beq
\label{4.23}
\sum_{i=1}^2\int \dd\mathbf{v} \frac{1}{2}m_i V^2
\left(
\begin{array}{cccc}
\mathbf{A}_{i}\\
\mathbf{B}_{i}\\
D_i\\
\mathbf{E}_{i}
\end{array}
\right)
=
\left(
\begin{array}{ccccc}
\mathbf{0}\\
\mathbf{0}\\
0\\
\mathbf{0}
\end{array}
\right),
\eeq
and
\beq
\label{4.23.1}
\sum_{i=1}^2\int \dd\mathbf{v} \frac{1}{2}m_i V^2 C_{i,\lambda\beta}\left(\partial_\lambda U_\beta+\partial_\beta U_\lambda-\frac{2}{d}\delta_{\lambda\beta}\nabla\cdot\mathbf{U}\right)=0.
\eeq
Since $\boldsymbol{\mathcal{A}}_i \propto \mathbf{A}_i$, $\boldsymbol{\mathcal{B}}_{ij} \propto \mathbf{B}_{ij}$, $\mathcal{C}_{i,\lambda\beta}
\propto C_{i,\lambda\beta}$, $\mathcal{D}_i \propto D_i$, and $\boldsymbol{\mathcal{E}}_i \propto \mathbf{E}_i$, then the solubility conditions \eqref{4.5.3} and \eqref{4.5.3.1} are fulfilled and so, there exist solutions to the integral equations \eqref{4.16}--\eqref{4.19}; this is the so-called Fredholm alternative \cite{MM56}.

\section{Navier--Stokes transport coefficients}
\label{sec5}

The forms of the constitutive equations for the irreversible fluxes to first-order in spatial gradients can be written using simple symmetry arguments \cite{G19}. While the mass flux $\mathbf{j}_i^{(1)}$ of the component $i$ is given by Eq.\ \eqref{5.1}, the pressure tensor $P_{\lambda\beta}$ has the form
\beq
\label{5.2}
P_{\lambda\beta}^{(1)}=-\eta \left(\partial_\lambda U_\beta+\partial_\beta U_\lambda-\frac{2}{d}\delta_{\lambda\beta}\nabla\cdot\mathbf{U}\right)-\delta_{\lambda\beta}\eta_\text{b}\nabla \cdot \mathbf{U},
\eeq
while the heat flux $\mathbf{q}^{(1)}$ can be written as
\beq
\label{5.3}
\mathbf{q}^{(1)}=-\sum_{i=1}^2\sum_{j=1}^2\; T^2 D_{q,ij} \nabla \ln n_j-T\kappa \nabla \ln T+\kappa_U\Delta\mathbf{U}.
\eeq
In Eqs.\ \eqref{5.2}--\eqref{5.3}, $\eta$ is the shear viscosity coefficient, $\eta_\text{b}$ is the bulk viscosity coefficient, $\kappa$ is the thermal conductivity coefficient, $\kappa_U$ is the velocity conductivity, and $D_{q,ij}$ are the partial contributions to the Dufour coefficients $D_{q,i}$ defined as \cite{G19}
\beq
\label{5.4}
D_{q,i}=\sum_{\ell=1}^2 D_{q,\ell i}.
\eeq

The transport coefficients associated with the mass flux are defined as
\beq
\label{5.5}
D_i^T=-\frac{m_i}{d\rho}\int\; \dd \mathbf{v}\; \mathbf{V}\cdot \boldsymbol{\mathcal{A}}_{i}(\mathbf{V}),
\eeq
\beq
\label{5.6}
D_{ij}=-\frac{\rho}{d\rho_j}\int\; \dd \mathbf{v}\; \mathbf{V}\cdot \boldsymbol{\mathcal{B}}_{ij}(\mathbf{V}),
\eeq
\beq
\label{5.6a}
D^U_i=-\frac{m_i}{d}\int\; \dd \mathbf{v}\; \mathbf{V}\cdot \boldsymbol{\mathcal{E}}_i(\mathbf{V}).
\eeq

As said in Sec.\ \ref{sec2}, in contrast to the mass flux, the pressure tensor and heat flux have kinetic and collisional contributions. To first-order, their kinetic contributions are
\beq
\label{5.7}
P_{\lambda\beta}^{\text{k}(1)}=\sum_{i=1}^2\int\; \dd \mathbf{v}\; m_i V_\lambda V_\beta f_i(\mathbf{V}),
\eeq
\beq
\label{5.8}
\mathbf{q}^{\text{k}(1)}=\sum_{i=1}^2\int\; \dd \mathbf{v}\; \frac{m_i}{2} V^2 \mathbf{V} f_i(\mathbf{V}).
\eeq
According to Eqs.\ \eqref{5.2} and \eqref{5.7}, the kinetic contribution $\eta_\text{k}$ to the shear viscosity can be written as  $\eta_\text{k}=\sum_{i=1}^2\eta_i^{\text{k}}$, where \cite{G19}
\beq
\label{5.9}
\eta_i^{\text{k}}=-\frac{1}{(d-1)(d+2)}\int\; \dd \mathbf{v}\; m_i V_\lambda V_\beta \mathcal{C}_{i,\lambda\beta}(\mathbf{V}).
\eeq
In the case of the heat flux, according to Eqs.\ \eqref{5.3} and \eqref{5.8}, the kinetic contribution $D_{q,ij}^\text{k}$  to the Dufour coefficient is
\beq
\label{5.10}
D_{q,ij}^{\text{k}}=-\frac{1}{d T^2}\int \dd\mathbf{v}\frac{1}{2} m_i
V^{2}\mathbf{V}\cdot \boldsymbol{\mathcal{B}}_{ij}\left( \mathbf{V}\right),
\eeq
while the kinetic contributions $\kappa_\text{k}$ and $\kappa^\text{k}_U$ to the thermal and velocity conductivity coefficients, respectively, can be written as $\kappa_\text{k}=\sum_{i=1}^2\kappa_i^{\text{k}}$ and $\kappa^\text{k}_U=\sum_{i=1}^2\kappa_i^{U \text{k}}$, where
\beq
\label{5.11}
\kappa_i^{\text{k}}=-\frac{1}{d T}\int
\dd\mathbf{v}\frac{1}{2}m_iV^2\mathbf{V}\cdot \boldsymbol{\mathcal{A}}_{i}\left(\mathbf{V}\right).
\eeq
\beq
\label{5.11a}
\kappa_i^{U \text{k}}=-\frac{1}{d}\int
\dd\mathbf{v}\frac{1}{2}m_iV^2\mathbf{V}\cdot \boldsymbol{\mathcal{E}}_{i}\left(\mathbf{V}\right).
\eeq

Collisional contributions to the pressure tensor and heat flux can be obtained by expanding Eqs.\ \eqref{2.22} and \eqref{2.23} to first-order in spatial gradients. A careful analysis shows that those collisional contributions are formally the same as those obtained in the dry granular case \cite{G19,GGG19b,GDH07,GHD07}. In particular, the bulk viscosity (which has only collisional contributions) can be written as
\beq
\label{5.11b}
\eta_\text{b}=\eta_\text{b}'+\eta_\text{b}'',
\eeq
where
\beqa
\label{5.12}
\eta_\text{b}^{\prime}&=&\frac{\pi^{(d-1)/2}}{\Gamma\left(\frac{d+3}{2}\right)}\frac{d+1}{2d^{2}}\sum_{i=1}^{2}\sum_{j=1}^{2}m_{ij}
\left(1+\alpha_{ij}\right)\chi_{ij}^{(0)}\sigma_{ij}^{d+1}\nonumber\\
& & \times \int \dd\mathbf{v}_1\int \dd\mathbf{v}_{2}f_{i}^{(0)}(\mathbf{V}_{1})f_{j}^{(0)}(\mathbf{V}_{2})g_{12},
\eeqa
and
\beq
\label{5.12.1}
\eta_\text{b}''=-\frac{\pi^{d/2}}{d\Gamma\left(\frac{d}{2}\right)}\sum_{i=1}^{2}\sum_{j=1}^{2}\mu_{ji}
\left(1+\alpha_{ij}\right)\chi_{ij}^{(0)} n_i n_j\sigma_{ij}^{d}\varpi_i.
\eeq
The second contribution $\eta_\text{b}''$ to $\eta_\text{b}$ was neglected in the previous works on granular mixtures   \cite{GDH07,GHD07,G19} because it was implicitly assumed that its contribution to the bulk viscosity was quite small. On the other hand, this influence was already accounted for in the pioneering studies on ordinary (elastic collisions) hard-sphere mixtures \cite{KS79a,KS79b,LCK83} and has been recently calculated \cite{GGG19b} in the case of (dry) polydisperse dense granular mixtures.

The collisional contribution $\eta_\text{c}$ to the shear viscosity is
\beqa
\label{5.13}
\eta^{\text{c}}&=&\frac{2 \pi^{d/2}}{d(d+2)\Gamma\left(\frac{d}{2}\right)}\sum_{i=1}^{2}\sum_{j=1}^{2}\mu_{ij}\left(1+\alpha_{ij}\right)
\chi_{ij}^{(0)}n_{i}\sigma_{ij}^{d}\eta_{j}^{\text{k}}\nonumber\\
& & +\frac{d}{d+2}\eta_\text{b}'.
\eeqa
The expressions of the collisional contributions to the heat flux transport coefficients are more intricate than that of $\eta_\text{b}$ and $\eta_\text{c}$. Their explicit forms can be found in Ref.\ \cite{G19}.

\section{Approximate results. Leading Sonine approximations}
\label{sec6}

The evaluation of the complete set of transport coefficients of the binary granular suspension is a quite long task. In this paper, we will focus on our attention in obtaining the transport coefficients associated with the mass flux ($D_{ij}$, $D_i^T$, and $D_i^U$), the shear viscosity coefficient $\eta$, and the partial temperatures $T_i^{(1)}$. To determine them, one has to solve the set of coupled linear integral equations \eqref{4.16}--\eqref{4.19.1} as well as to know the forms of the zeroth-order distributions $f_i^{(0)}$. With respect to the latter, as noted in Sec.\ \ref{sec3}, $f_i^{(0)}$ is well represented by the Maxwellian velocity distribution function
\beq
\label{5.18}
f_i^{(0)}(\mathbf{V})\to f_{i,\text{M}}(\mathbf{V})=n_i \left(\frac{m_i}{2\pi T_i^{(0)}}\right)^{d/2}\exp\left(-\frac{m_i V^2}{2T_i^{(0)}}\right).
\eeq
This means that we neglect here non-Gaussian corrections to the distributions $f_i^{(0)}$ and hence, one expects to get simple but accurate expressions for the transport coefficients. By using the Maxwellian approximation \eqref{5.18}, the collisional contribution $\eta_\text{b}'$ is  \beqa
\label{5.18.1}
\eta_\text{b}^{\prime}&=&\frac{\pi^{(d-1)/2}}{d^2\Gamma\left(\frac{d}{2}\right)}v_0\sum_{i=1}^{2}\sum_{j=1}^{2}m_{ij}
\left(1+\alpha_{ij}\right)\chi_{ij}^{(0)}n_i n_j \sigma_{ij}^{d+1}\nonumber\\
& & \times \left(\frac{\beta_i+\beta_j}{\beta_i\beta_j}\right)^{1/2}.
\eeqa

Regarding the unknowns $\left(\boldsymbol{\mathcal{A}}_{i}, \boldsymbol{\mathcal{B}}_{ij}, \mathcal{C}_{i,\lambda\beta}, \mathcal{D}_i, \boldsymbol{\mathcal{E}}_{i}\right)$, as usual we will expand them in a series expansion of orthogonal polynomials (Sonine polynomials) \cite{BP04} and we will truncate this expansion by considering only the leading term (lowest degree polynomial). In particular, the collisional contribution $\eta_\text{b}^{\prime\prime}$ will be estimated latter when we determine $\varpi_i$ in the first Sonine approximation.

\subsection{Diffusion transport coefficients}
\label{sub6.1}

In the case of the transport coefficients $D_{ij}$, $D_i^T$, and $D_i^U$, the leading Sonine approximations to $\boldsymbol{\mathcal{A}}_{i}$, $\boldsymbol{\mathcal{B}}_{ij}$, and $\boldsymbol{\mathcal{E}}_{i}$ are, respectively,
\beq
\label{5.16}
\boldsymbol{\mathcal{A}}_i(\mathbf{V})\rightarrow-\frac{\rho}{n_iT_i}D_i^Tf_{i,\text{M}}(\mathbf{V})\mathbf{V},
\eeq
\beq
\label{5.17}
\boldsymbol{\mathcal{B}}_{ij}(\mathbf{V})\rightarrow-\frac{m_i\rho_j}{\rho n_iT_i}D_{ij}f_{i,\text{M}}(\mathbf{V})\mathbf{V},
\eeq
\beq
\label{5.17a}
\boldsymbol{\mathcal{E}}_i(\mathbf{V})\rightarrow-\frac{D_i^U}{n_iT_i}f_{i,\text{M}}(\mathbf{V})\mathbf{V}.
\eeq

In order to determine the above diffusion coefficients, we substitute first $\boldsymbol{\mathcal{A}}_{i}$, $\boldsymbol{\mathcal{B}}_{ij}$, and $\boldsymbol{\mathcal{E}}_{i}$  by their leading Sonine approximations \eqref{5.16}--\eqref{5.17a} in Eqs. \eqref{4.16}, \eqref{4.17}, and \eqref{4.19.1}, respectively. Then, we multiply these equations by $m_i \mathbf{V}$ and integrate over velocity. The final forms of the set of algebraic equations defining the transport coefficients $D_i^T$, $D_{ij}$, and $D_i^U$ are given by Eqs.\ \eqref{5.19}--\eqref{5.21} of  the Appendix \ref{appC}.

The solution to the set of Eqs.\ \eqref{5.19}--\eqref{5.21} provides the dependence of the (relevant) diffusion coefficients $D_{11}$, $D_{12}$, $D_1^T$, and $D_1^U$ on the coefficients of restitution $\al_{ij}$, the concentration $x_1$, the solid volume fraction $\phi$, the masses and diameters of the constituents of the mixture, and the (reduced) background temperature $T_\text{ex}^*$. In particular, the expression of $D_1^U$ is
\beq
\label{5.21.3}
D_1^U=\rho_1\rho_2 \frac{\gamma_1-\gamma_2}{\rho \nu_{D}+\rho_1\gamma_2+\rho_2\gamma_1},
\eeq
where $\nu_D$ is defined by Eq.\ \eqref{5.21.5}. The explicit form of the thermal diffusion coefficient $D_1^T$ is given by Eq.\ \eqref{5.21.4}. The expressions of $D_{11}$ and $D_{12}$ can be obtained by solving the set of Eqs.\ \eqref{5.20}. Their forms are very large and will be omitted here for the sake of simplicity.

Equations \eqref{5.21.3} and \eqref{5.21.4} show that $D_1^U$ and $D_1^T$ are antisymmetric with respect to the change $1\leftrightarrow 2$ as expected. This can be easily verified since $x_1\tau_1+x_2\tau_2=1$, $\Delta_{\theta,1}=-(x_2/x_1)\Delta_{\theta,2}$, and
\beq
\label{5.21.6}
\frac{\partial p^*}{\partial \theta}=\frac{\pi^{d/2}}{d\Gamma\left(\frac{d}{2}\right)}\sum_{i=1}^2\sum_{j=1}^2\mu_{ji}x_in_j \sigma_{ij}^d \chi_{ij}^{(0)} \Delta_{\theta,i}(1+\al_{ij}),
\eeq
where $p^*\equiv p/(n T)$ is the reduced hydrostatic pressure. Furthermore, in the case of mechanically equivalent particles ($m_1=m_2$, $\sigma_1=\sigma_2$, $\chi_{ij}^{(0)}=\chi^{(0)}$, and $\al_{ij}=\al$), Eqs.\ \eqref{5.20} and \eqref{5.21.4} yield $x_1 D_{11}^*+x_2D_{12}^*=0$ and $D_1^{T*}=0$, as expected. Here, we have introduced the scaled coefficients $D_{ij}^*\equiv D_{ij}(\al)/D_{ij}(1)$ and $D_{1}^{T*}\equiv D_{1}^T(\al)/D_{1}^T(1)$ where $D_{ij}(1)$ and $D_1^T(1)$ refer to the values of $D_{ij}$ and $D_1^T$, respectively, for elastic collisions. The above relations confirm the self-consistency of the expressions for the diffusion coefficients reported in this paper.

\subsection{Shear viscosity coefficient}
\label{sub6.2}

The kinetic contribution to the shear viscosity $\eta_\text{k}=\eta_1^{\text{k}}+\eta_2^{\text{k}}$, where the partial contributions $\eta_i^{\text{k}}$ are defined by Eq.\ \eqref{5.9}. To determine the kinetic coefficients $\eta_i^{\text{k}}$, the function $\mathcal{C}_{i,\lambda\beta}(\mathbf{V})$ is estimated by its leading Sonine approximation
\beq
\label{5.22}
\mathcal{C}_{i,\lambda\beta}(\mathbf{V}) \rightarrow -f_{i,\text{M}}(\mathbf{V}) R_{i,\lambda\beta}(\mathbf{V})\frac{\eta_i^{\text{k}}}{n_i{T_i^{(0)}}^2},
\eeq
where
\beq
\label{5.23}
 R_{i,\lambda\beta}(\mathbf{V})=m_i\left(V_\lambda V_\beta-\frac{1}{d}\delta_{\lambda\beta}V^2\right).
\eeq

As in the case of the diffusion coefficients, the partial contributions $\eta_i^{\text{k}}$ are obtained by substituting Eq.\ \eqref{5.22} into the integral equation \eqref{4.18}, multiplying it by $R_{i,\lambda\beta}$ and integrating over the velocity. After some algebra, one achieves the set of algebraic equations \eqref{5.24}. The solution to the set \eqref{5.24} provides the partial contributions $\eta_i^\text{k}$. Their sum then gives the kinetic coefficient $\eta_\text{k}$. Finally, by adding this to the collisional contribution \eqref{5.13}  we have the total shear viscosity.

\subsection{First-order contributions to the partial temperatures}	
\label{sub6.3}

Finally, we consider the first-order contribution $T_i^{(1)}$ to the partial temperature $T_i$. This coefficient is defined by Eqs.\ \eqref{4.20T1} and \eqref{4.20T2}. As said before, the coefficients $T_i^{(1)}$ $(i=1,2)$ have been recently determined for dry granular mixtures \cite{GGG19b}. To determine $\varpi_i$, we consider the leading Sonine approximation to $\mathcal{D}_i(\mathbf{V})$ given by
\beq
\label{5.29}
\mathcal{D}_i(\mathbf{V})\rightarrow f_{i\text{M}}(\mathbf{V})W_i(\mathbf{V})\frac{\varpi_i}{T_i^{(0)}},\quad
W_i(\mathbf{V})=\frac{m_iV^2}{2T_i^{(0)}}-\frac{d}{2}.
\eeq

The coefficients $\varpi_i$ are coupled with the coefficients $\zeta^{(1,1)}$ through Eq.\ \eqref{b16}. The explicit relation between $\zeta^{(1,1)}$ and $\varpi_i$ can be easily obtained by substitution of Eq.\ \eqref{5.29} into Eq.\ \eqref{b16}, with the result
\beq
\label{5.30.1}
\zeta^{(1,1)}=\sum_{i=1}^2\; \xi_i \varpi_i,
\eeq
where
\beqa
\label{5.30.2}
\xi_i&=&\frac{3\pi^{(d-1)/2}}{2d\Gamma\left(\frac{d}{2}\right)}\frac{v_0^3}{n T T_i^{(0)}}\sum_{j=1}^2 n_i n_j \sigma_{ij}^{d-1}
\chi_{ij}^{(0)}m_{ij}(1-\al_{ij}^2)\nonumber\\
& & \times \left(\beta_i+\beta_j\right)^{1/2}\beta_i^{-3/2}\beta_j^{-1/2}.
\eeqa

As usual, in order to obtain the coefficients $\varpi_i$, one substitutes first  Eq.\ \eqref{5.29} into Eq.\ \eqref{4.19} and then multiplies it with $m_iV^2$ and integrates over the velocity. After some algebra, one gets the set of coupled equations \eqref{5.31}. A careful inspection to the set of Eqs.\ \eqref{5.31} shows that $\varpi_1=-(x_2/x_1) \varpi_2$ as the solubility condition \eqref{4.5.3.1} requires. This is because $x_1\tau_1+x_2\tau_2=1$, $\Delta_{\theta,2}=-(x_2/x_1)\Delta_{\theta,1}$, and $\omega_{11}-(x_1/x_2)\omega_{12}+\xi_1/x_1=\omega_{22}-(x_2/x_1)\omega_{21}+\xi_2/x_2$. The condition $x_1\varpi_1+x_2\varpi_2=0$ guarantees that the temperature $T$ is not affected by the spatial gradients.

The solution to Eq.\ \eqref{5.31} gives $\varpi_1$ in terms of the parameters of the mixture. On the other hand, its explicit form is relatively long and is omitted here for the sake of brevity. A simple but interesting case corresponds to elastic collisions (molecular or ordinary suspensions) where $\xi_i=0$, $\tau_i=1$, $\beta_1=2\mu_{12}$, $\beta_2=2\mu_{21}$,  $\Delta_{\theta,i}=\Delta_{x_1,i}=\Delta_{\lambda_1,i}=\Delta_{\phi,i}=0$, and so $\varpi_1$ is simply given by
\begin{widetext}
\beq
\label{5.35.1}
\varpi_1=\frac{4\pi^{d/2}}{d^2\Gamma\left(\frac{d}{2}\right)}T\frac{n_2\sigma_{12}^d\chi_{12}^{(0)}\left(x_2\mu_{21}-x_1\mu_{12}\right)
+\frac{1}{2}x_2\left(n_1\sigma_1^d \chi_{11}^{(0)}-n_2\sigma_2^d \chi_{22}^{(0)}\right)}{\omega_{11}-\frac{x_1}{x_2}\omega_{12}-
2\left(x_2\gamma_1+x_1\gamma_2\right)}.
\eeq
\end{widetext}
Equation \eqref{5.35.1} is consistent with the one derived many years ago by Karkheck and Stell \cite{KS79b} for ordinary hard-sphere mixtures ($\gamma_1=\gamma_2=0$).

Once the first-order contributions to the partial temperatures are known, the first-order contribution $\zeta_U$ to the cooling rate can be explicitly obtained from Eqs.\ \eqref{4.20.1}, \eqref{b8}, and \eqref{5.30.1}. In addition, the contribution $\eta_\text{b}''$ to the bulk viscosity $\eta_\text{b}$ can be obtained from Eq.\ \eqref{5.12.1} and hence, the bulk viscosity is completely determined by Eqs.\ \eqref{5.12.1} and \eqref{5.18.1}.

\section{Some illustrative systems}
\label{sec7}

\begin{figure}
\includegraphics[width=0.8\columnwidth]{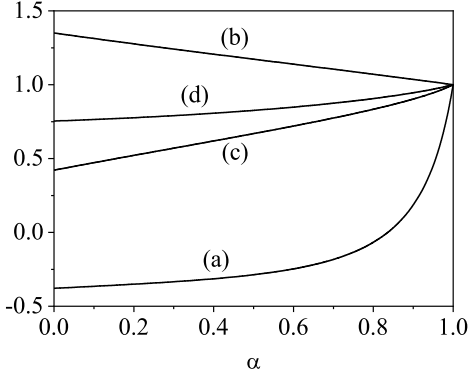}
\caption{Plot of the (reduced) diffusion coefficients $D_1^{T*}$ (a), $D_1^{U*}$ (b), $D_{11}^{*}$ (c), and $D_{12}^{*}$ (d) as a function of the common coefficient of restitution $\al$ for an equimolar mixture ($x_1=\frac{1}{2}$) of hard spheres ($d=3$) with $\sigma_1/\sigma_2=1$, $m_1/m_2=4$, $\phi=0.2$, and $T_\text{ex}^*=0.1$.}
\label{fig5}
\end{figure}

The results derived in Sec.\ \ref{sec6} for the diffusion transport coefficients, the shear and bulk viscosities and the first-order contributions to the partial temperatures and the cooling rate depend on the background temperature $T_\text{ex}$, the concentration $x_1$, the density or volume fraction $\phi$, and the mechanical parameters of the binary mixture (masses, diameters, and coefficients of restitution). As in our previous paper \cite{KG13} on dilute granular suspensions, given that the new relevant feature is the dependence of the transport coefficients on inelasticity, we scale these coefficients with respect to their values for elastic collisions. Thus, the scaled transport coefficients depend on the parameter space: $\left\{T_\text{ex}^*, x_1, m_1/m_2, \sigma_1/\sigma_2, \phi, \al_{11}, \al_{22}, \al_{12}\right\}$. Moreover, for the sake of simplicity, the case of a common coefficient of restitution ($\al_{11}=\al_{22}=\al_{12}\equiv \al$) of an equimolar hard-sphere mixture ($x_1=\frac{1}{2}$ and $d=3$) with a background temperature $T_\text{ex}^*=0.1$ is considered. This reduces the parameter space to four quantities: $\left\{m_1/m_2, \sigma_1/\sigma_2, \phi, \al\right\}$.

To display the dependence of the transport coefficients on the coefficient of restitution, we have to provide the form for the pair distribution function $\chi_{ij}^{(0)}$. In the case of spheres ($d=3$), a good approximation of $\chi_{ij}^{(0)}$ is \cite{B70,GH72,LL73}
\beq
\label{7.1}
\chi_{ij}=\frac{1}{1-\phi}+\frac{3}{2}\frac{\phi}{(1-\phi)^2}\frac{\sigma_i \sigma_j M_2}{\sigma_{ij}M_3}+\frac{1}{2}\frac{\phi^2}{(1-\phi)^3}
\left(\frac{\sigma_i \sigma_j M_2}{\sigma_{ij}M_3}\right)^2,
\eeq
where $M_\ell=\sum_i x_i \sigma_i^\ell$. In addition, the functions $R_i$ are defined by Eq.\ \eqref{2.5.3}.

\begin{figure}
\includegraphics[width=0.8\columnwidth]{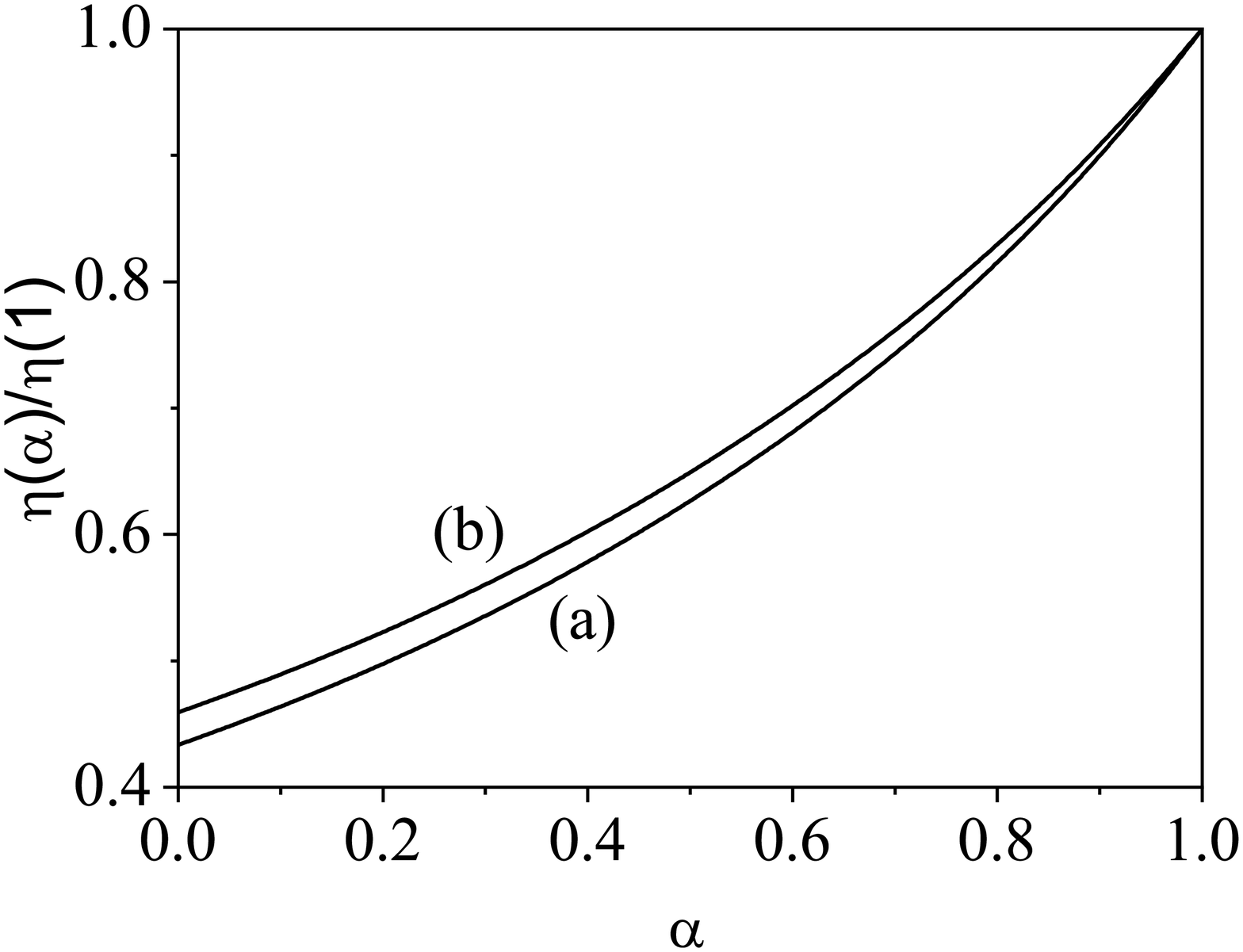}
\caption{Plot of the (reduced) shear viscosity coefficient $\eta(\al)/\eta(1)$ as a function of the common coefficient of restitution $\al$ for an equimolar mixture ($x_1=\frac{1}{2}$) of hard spheres ($d=3$) with $\sigma_1/\sigma_2=1$, $\phi=0.2$, and $T_\text{ex}^*=0.1$. Two different values of the mass ratio are considered: $m_1/m_2=0.5$ (a) and $m_1/m_2=4$ (b).}
\label{fig6}
\end{figure}

\begin{figure}
\includegraphics[width=0.8\columnwidth]{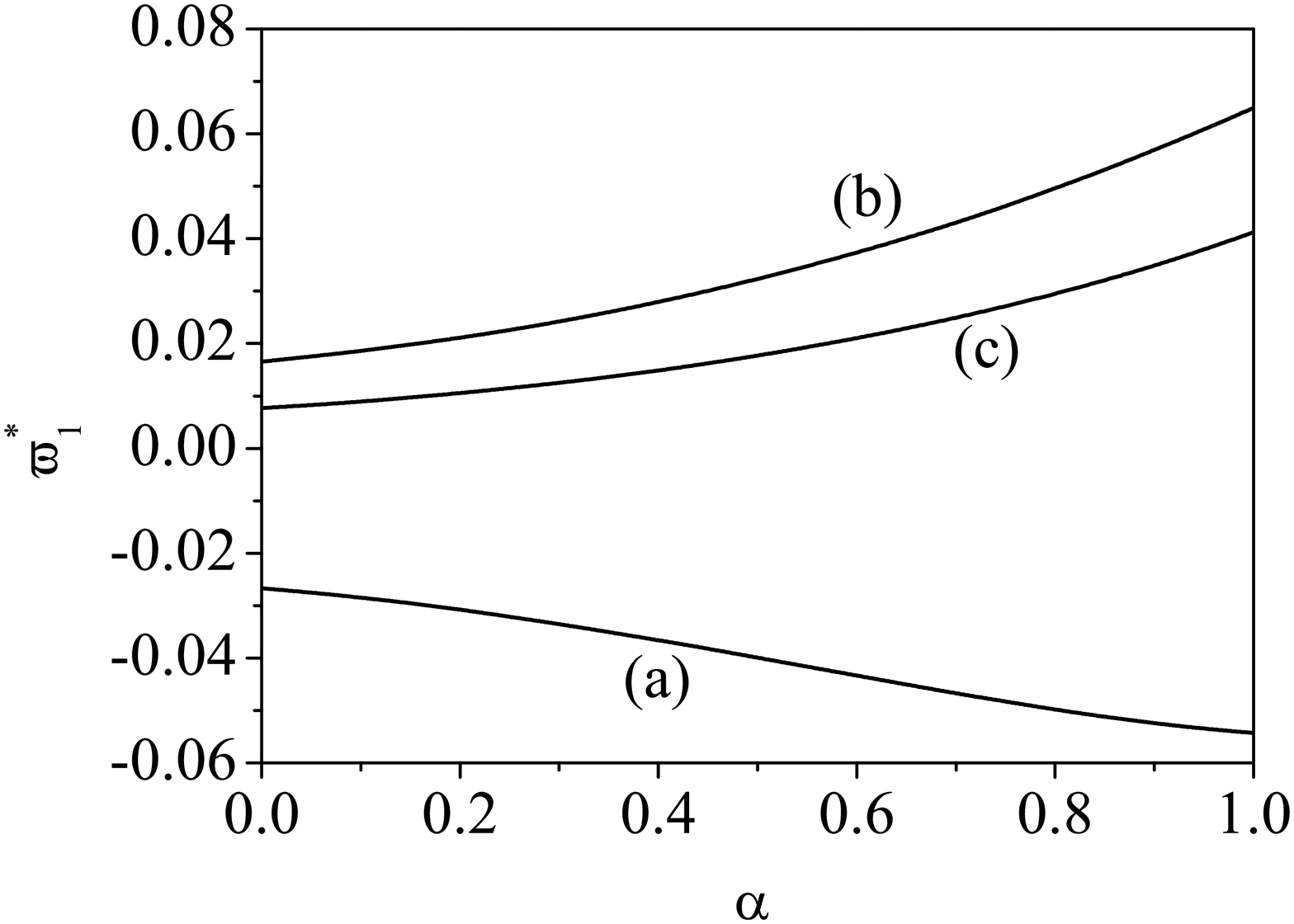}
\caption{Plot of the (reduced) coefficient $\varpi_1^*$ as a function of the common coefficient of restitution $\alpha$ for an equimolar mixture ($x_1=\frac{1}{2}$) of hard spheres ($d=3$) with $\sigma_1/\sigma_2=1$, $\phi=0.2$, and $T_\text{ex}^*=0.1$. Three different values of the mass ratio are considered: $m_1/m_2=0.5$ (a), $m_1/m_2=4$ (b), and $m_1/m_2=10$ (c).}
\label{fig7}
\end{figure}
\begin{figure}
\includegraphics[width=0.8\columnwidth]{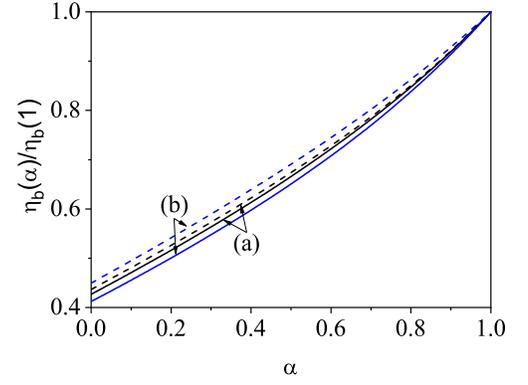}
\caption{Plot of the (reduced) bulk viscosity coefficient $\eta_\text{b}(\al)/\eta_\text{b}(1)$ as a function of the common coefficient of restitution $\al$ for an equimolar mixture ($x_1=\frac{1}{2}$) of hard spheres ($d=3$) with $\sigma_1/\sigma_2=1$, $\phi=0.2$, and $T_\text{ex}^*=0.1$. Two different values of the mass ratio are considered: $m_1/m_2=0.5$ (a) and $m_1/m_2=10$ (b). The dashed lines are the results for the (reduced) bulk viscosity when the contribution $\eta_\text{b}''$ to $\eta_\text{b}$ is neglected.}
\label{fig8}
\end{figure}
\begin{figure}
\includegraphics[width=0.8\columnwidth]{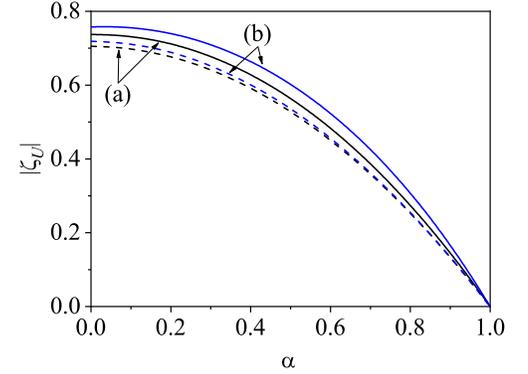}
\caption{Plot of the magnitude of the (reduced) coefficient $\zeta_U$ as a function of the common coefficient of restitution $\alpha$ for an equimolar mixture ($x_1=\frac{1}{2}$) of hard spheres ($d=3$) with $\sigma_1/\sigma_2=1$, $\phi=0.2$, and $T_\text{ex}^*=0.1$. Two different values of the mass ratio are considered: $m_1/m_2=0.5$ (a) and $m_1/m_2=10$ (b). The dashed lines are the results for the coefficient $\zeta_U$ when the contribution $\zeta^{(1,1)}$ to $\zeta_U$ is neglected.}
\label{fig9}
\end{figure}

Figure \ref{fig5} shows the $\al$-dependence of the reduced diffusion coefficients $D_{ij}^*$, $D_{1}^{T*}$, and $D_1^{U*}$ for $m_1/m_2=4$, $\sigma_1/\sigma_2=1$, and $\phi=0.1$. We recall that $D_{ij}^*\equiv D_{ij}(\al)/D_{ij}(1)$, $D_{1}^{T*}\equiv D_{1}^T(\al)/D_{1}^T(1)$, and $D_1^{U*}\equiv D_1^U(\al)/D_1^U(1)$, where $D_{ij}(1)$, $D_1^U(1)$, and $D_1^T(1)$ are the values of the diffusion transport coefficients for elastic collisions. It is quite apparent first that the effect of inelasticity on diffusion coefficients is in general significant since the forms of the scaled coefficients $D_{ij}^*$, $D_1^{U*}$, and $D_1^{T*}$ differs clearly from their elastic counterparts. This is specially relevant in the case of the thermal diffusion coefficient $D_1^{T*}$.  In addition, a comparison with the results obtained for dry granular mixtures (see for instance, Figs.\ 5.5, 5.6, and 5.7 of Ref.\ \cite{G19} for the same mixture parameters) reveals significant differences between dry (no gas phase) and gas-solid flows when grains are mechanically different. Thus, while $D_{11}^*$ and $D_{12}^{*}$ increase with inelasticity for dry granular mixtures, the opposite happens for granular suspensions since they decrease as increasing inelasticity. The qualitative $\al$-dependence of $D_1^{T*}$ is similar in both dry and gas-solid flows, although the influence of inelasticity on $D_1^{T*}$ is much more important in the latter case.

We consider now the (reduced) shear viscosity $\eta^*\equiv \eta(\al)/\eta(1)$. Figure \ref{fig6} shows $\eta^*$ versus $\al$ for $\sigma_1/\sigma_2=1$, $\phi=0.2$, and two different values of the mass ratio. As with the diffusion coefficients, the effect of inelasticity on the shear viscosity is again significant since the inelasticity hinders the transport of momentum. Regarding the comparison with dry granular mixtures, we find qualitative differences since both theory \cite{GD99a,L05,G13,MGH12} and simulations \cite{G19,MSG05} have shown that while for relatively dilute dry granular gases $\eta^*$ increases with inelasticity, the opposite occurs for sufficiently dense granular mixtures. This non-monotonic behavior with density contrasts with the results obtained here for multicomponent granular suspensions since the scaled coefficient $\eta^*$ always decreases with increasing inelasticity regardless of the value of the solid volume fraction $\phi$. With respect to the influence of the mass ratio on the shear viscosity, we see that its impact on $\eta^*$ is relatively small. In particular, at a given value of $\al$, $\eta^*$ decreases with decreasing the mass ratio $m_1/m_2$.

An interesting quantity is the first-order contribution $\varpi_1$ to the partial temperature $T_1$. The reduced coefficient $\varpi_1^*\equiv (n\sigma_{12}^2 v_0/T)\varpi_1$ is plotted in Fig.\ \ref{fig8} as a function of $\al$ for $\sigma_1/\sigma_2=1$, $\phi=0.2$, and three different values of the mass ratio. We observe that the influence of inelasticity on $\varpi_1^*$ is important, specially for high mass ratios. However, Fig.\ \ref{fig8} highlights that the magnitude of $\varpi_1^*$ is much smaller than the other transport coefficients and hence, the impact of the first-order contribution $T_1^{(1)}$ on both the bulk viscosity $\eta_\text{b}$ (through the coefficient $\eta_b''$) and the first-order contribution $\zeta_U$ (through the coefficient $\zeta^{(1,1)}$) to the cooling rate is expected to be small. This is confirmed by Figs.\ \ref{fig8} and \ref{fig9} for the reduced coefficients $\eta_\text{b}(\al)/\eta_\text{b}(1)$ and $\zeta_U$, respectively. It is quite apparent that the theoretical predictions for the above coefficients with and without the contribution of $\varpi_1^*$ are practically identical, specially in the case of the bulk viscosity. As the shear viscosity coefficient, we also see that the bulk viscosity decreases with increasing inelasticity (independent of the mass ratio considered). Moreover, as for dry granular mixtures \cite{GGG19b}, the coefficient $\zeta_U$ is always negative and its magnitude increases with inelasticity.

In summary, the mass and momentum transport coefficients for a multicomponent granular suspension differ significantly from those for dry granular mixtures. In most of cases, the differences become greater with increasing inelasticity, and depending on the cases, there is a relevant influence of the mass ratio.

\section{Discussion}
\label{sec8}

This paper has been focused on the determination of the Navier--Stokes transport coefficients of a binary granular suspension at moderate densities. The starting point of our study has been the set of Enskog kinetic equations for the velocity distribution functions $f_i(\mathbf{r}, \mathbf{v},t)$ of the solid particles. The effect of the gas phase on the solid particles has been accounted for by an effective force constituted by two terms, namely a viscous drag force proportional to the velocity of the particles and a stochastic Langevin-like term. Therefore, we have considered a simplified model where the effect of the interstitial gas on grains is explicitly accounted for but the state of the surrounding gas is assumed to be independent of the solid particles. On the other hand, this model is inspired in numerical and experimental results that can be found in the granular literature \cite{YS09b}. This fact is reflected in the functional dependence of the friction coefficients $\gamma_i$ on both the partial $\phi_i$ and global $\phi=\phi_1+\phi_2$ volume fractions, and the mechanical properties of grains (masses $m_i$ and diameters $\sigma_i$).

We have derived the Navier--Stokes hydrodynamic equations in two steps. First, the macroscopic balance equations \eqref{2.15}--\eqref{2.17} have been obtained from the Enskog equation \eqref{2.1}. Particularly, these equations include terms that account for the impact of the gas phase on grains and the kinetic and collisional contributions to the fluxes are expressed as functionals of the velocity distribution functions $f_i$. Second, the mass, momentum, and heat fluxes, together with the cooling rate appearing in the hydrodynamic equations have been evaluated by solving the Enskog equation by means of the Chapman--Enskog method up to first order in the spatial gradients. The constitutive equation for the mass flux is given by Eq.\ \eqref{5.1} where the diffusion coefficients $D_i^T$, $D_{ij}$, and $D_i^U$ are defined by Eqs.\ \eqref{5.5}--\eqref{5.6a}, respectively. The pressure tensor is given by Eq.\ \eqref{5.2} where the bulk viscosity $\eta_\text{b}$ is defined by Eq.\ \eqref{5.11b} and the shear viscosity $\eta$ is defined by Eqs.\ \eqref{5.9} (kinetic contribution) and \eqref{5.13} (collisional contribution). Finally, the constitutive equation for the heat flux is given by Eq.\  \eqref{5.3} where the kinetic contributions to the Dufour coefficients $D_{q,ij}$, the thermal conductivity $\kappa$, and the velocity conductivity $\kappa_U$ are given by Eqs.\ \eqref{5.10}, \eqref{5.11}, and \eqref{5.11a}, respectively. Within the context of small gradients, all the above results apply in principle to arbitrary degree of inelasticity and are not restricted to specific values of the parameters of the mixture. The present work extends to moderate densities a previous analysis carried out for dilute bidisperse granular suspensions \cite{KG13,KG19}.

Before considering inhomogeneous situations, the homogeneous steady state has been studied. In this state, the distributions $f_{i,s}$ adopt the form \eqref{3.12} where the scaled distributions $\varphi_{i,s}$ depend on the steady temperature $T_s$ through the dimensionless velocity $\mathbf{c}=\mathbf{v}/v_0(T_s)$ and the (scaled) temperature $\theta=T_s/T_\text{ex}$. This scaling differs from the one assumed for dry granular mixtures \cite{GD99b} where the temperature dependence of $\varphi_i$ is only encoded through the dimensionless velocity $\mathbf{c}$. Although the exact form of the distributions $\varphi_{i,s}$ is not known, in order to estimate the partial temperatures $T_{i,s}/T_\text{ex}$, the distributions $\varphi_{i,s}$ have been approximated by the Maxwellian distributions \eqref{3.17}. This has allowed us to explicitly get the partial temperatures in terms of the parameters of the mixture. In spite of the crudeness of the above approximation, the theoretical predictions for $T_{1,s}/T_{2,s}$ agree well with MD simulations, specially for moderately dense systems. The goodness of the comparison supports the use of the Maxwellian approximation \eqref{3.17} in the evaluation of the transport coefficients. However, we find some discrepancies between theory and simulations that could be mitigated if one would consider the influence of the fourth cumulants on the distributions $\varphi_{i,s}$. We plan to calculate these cumulants in the near future and perform more simulations to assess the reliability of the Enskog theoretical predictions for homogeneous steady states.

Once the steady reference state is well characterized, the diffusion coefficients, the bulk and shear viscosities, and the first-order contributions to the partial temperatures and the cooling rate have been determined. As usual, in order to achieve explicit expressions for the above transport coefficients, the leading terms in a Sonine polynomial expansion have been considered. The explicit forms of the transport coefficients have been displayed along Sec.\ \ref{sec6} and the Appendix \ref{appC}: the coefficients $D_{11}$ and $D_{12}$ are the solutions of the algebraic equations \eqref{5.20}, the coefficients $D_1^U$ and $D_1^T$ are given by Eqs.\ \eqref{5.21.3} and \eqref{5.21.4}, respectively, the shear viscosity $\eta$ and the first-order coefficients $\varpi_i$ are the solutions of Eqs.\ \eqref{5.24} and \eqref{5.31}, respectively, and the first-order contribution $\zeta_U=\zeta^{(1,0)}+\zeta^{(1,1)}$ to the cooling rate is given by Eqs.\ \eqref{b8}, \eqref{5.30.1} and \eqref{5.30.2}. An interesting point is that these coefficients are defined not only in terms of the hydrodynamic fields in the steady state but, in addition, there are contributions to the transport coefficients coming from the derivatives of the temperature ratio in the vicinity of the steady state. These contributions can be seen as a measure of the departure of the perturbed state from the steady reference state. The inclusion of the above derivatives introduces conceptual and practical difficulties not present in the case of dry granular mixtures \cite{GDH07,GHD07}.

In reduced forms, the diffusion transport coefficients and the shear viscosity coefficient of the granular suspension exhibit a complex dependence on the parameter space of the problem. In particular, Fig.\ \ref{fig5} highlights the significant impact of the gas phase on the mass transport since the $\al$-dependence of the Navier--Stokes transport coefficients $D_1^T$, $D_1^U$, and $D_{ij}$ is in general different from the one found in the case of dry granular mixtures \cite{G19}. Regarding the shear viscosity coefficient $\eta$, a comparison with the dry granular results \cite{G19} shows a qualitative agreement between dry and granular suspensions for not quite high densities although important quantitative differences are found. Apart from these coefficients, the first-order contributions $\varpi_i$ to the partial temperatures $T_i$ have been also computed. The evaluation of these coefficients is interesting by itself and also because they are involved in the calculation of both the bulk viscosity $\eta_\text{b}$ and the first-order contribution $\zeta_U$ to the cooling rate. The results obtained here show that the magnitude of $\varpi_1$ is in general very small (in fact, much smaller than the one recently found \cite{GGG19b} in the absence of gas phase) and hence, its impact on $\eta_\text{b}$ and $\zeta_U$ is very tiny [see Figs.\ \ref{fig8} and \ref{fig9}]. This conclusion contrasts with recent findings for dry granular mixtures \cite{GGG19b} where the influence of $\varpi_1$ on both the bulk viscosity and the cooling rate must be taken into account for strong inelasticities and disparate masses.

In a subsequent paper, we plan to determine the heat flux transport coefficients and to perform a linear stability analysis of the homogeneous steady state as a possible application. In particular, given that the homogeneous steady state is stable in the dilute limit, we want to see if the density corrections to the transport coefficients can modify the stability of the above homogeneous state. In addition, it is also quite apparent that the reliability of the theoretical results derived here (which have been obtained under certain approximations) should be assessed against computer simulations. As happens for dry granular mixtures \cite{BRCG00,MG03,GM03,GM04,GV09,GV12,BR13,MGH14}, we expect that the present results stimulate the performance of appropriate simulations for bidisperse granular suspensions. In particular, we plan to undertake simulations to obtain the tracer diffusion coefficient (namely, a binary mixture where the concentration of one of the components is negligible) in a similar way as in the case of granular mixtures \cite{BRCG00,GV09,GV12}. Moreover, we also plan to carry out simulations to measure the Navier--Stokes shear viscosity $\eta$ by studying the decay of a small perturbation to the transversal component of the velocity field \cite{BRC99}. Another possible project for the next future is to consider inelastic rough hard spheres in order to assess the impact of friction on the transport properties of the granular suspension. Studies along these lines will be worked out in the near future.

\acknowledgments

The work of R.G.G. and V.G. has been supported by the Spanish Government through Grant No. FIS2016-76359-P and by the Junta de Extremadura
(Spain) Grant Nos. IB16013 and GR18079, partially financed by ``Fondo Europeo de Desarrollo Regional'' funds. The research of R.G.G. has been also supported by the predoctoral fellowship BES-2017-079725 from the Spanish Government.

\appendix
\section{Derivatives of the temperature ratio in the vicinity of the steady state}
\label{appA}

In this Appendix, the derivatives of the temperature ratio $\tau_1=T_1^{(0)}/T$ with respect to $\theta$, $\lambda_1$, $x_1$, and $\phi$  in the vicinity of the steady state are evaluated.

Let us consider first the derivative $(\partial \tau_1/\partial \theta)_{x_1,\lambda_1,\phi}$. To get it, we consider Eq.\ \eqref{4.13} for $i=1$:
\beq
\label{a1}
\Lambda^* \theta \frac{\partial \tau_1}{\partial \theta}=-\tau_1 \Lambda^*+\Lambda_1^*,
\eeq
where $\Lambda^*$ and $\Lambda_1^*$ are defined by Eqs.\ \eqref{4.13.1} and \eqref{4.14}, respectively. According to Eq.\ \eqref{3.18}, the (reduced) partial cooling rate $\zeta_{1,0}^*$ can be written as
\beq
\label{a2}
\zeta_{1,0}^*=\tau_1^{1/2}M_1^{-1/2}\zeta_1'(x_1,\beta),
\eeq
where $\beta=\beta_1/\beta_2=m_1\tau_2/(m_2\tau_1)$, $\tau_2=(1-x_1\tau_1)/x_2$, and
\beqa
\label{a3}
\zeta_1'(x_1,\beta)&=&\frac{\sqrt{2}\pi^{\left(d-1\right)/2}}{d\Gamma\left(\frac{d}{2}\right)}x_1\chi_{11}^{(0)}
\left(\frac{\sigma_{1}}{\sigma_{12}}\right)^{d-1}
\left(1-\alpha_{11}^2\right)\nonumber\\
& & +\frac{4\pi^{\left(d-1\right)/2}}{d\Gamma\left(\frac{d}{2}\right)}x_2\chi_{12}^{(0)}\mu_{21}\left(1+\beta\right)^{1/2}
\left(1+\alpha_{12}\right)\nonumber\\
& & \times \left[1-\frac{\mu_{21}}{2}\left(1+\alpha_{12}\right)\left(1+\beta\right)\right].
\eeqa
At the steady state, $\Lambda^*=\Lambda_1^*=\Lambda_2^*=0$ and hence, according to Eq.\ \eqref{a1}, the derivative $\partial \tau_1/\partial \theta$ becomes indeterminate. On the other hand, as for dilute multicomponent granular suspensions \cite{KG13}, the above problem can be fixed by applying l'H\^opital's rule. In this case, we take first the derivative with respect to $\theta$ in both sides of Eq. \eqref{a1} and then take the steady-state limit. After some algebra, one easily achieves the following quadratic equation for  the derivative $\Delta_{\theta,1}=\left(\partial\tau_1/\partial\theta\right)_{\text{s}}$:
\beqa
\label{a4}
& & \theta\Lambda_{1}^{(\theta)}\Delta_{\theta,1}^2+\left(\theta\Lambda_{0}^{(\theta)}+
\tau_{1}\Lambda_{1}^{(\theta)}-\Lambda_{11}^{(\theta)}\right)\Delta_{\theta,1}-
\Lambda_{10}^{(\theta)}\nonumber\\
& & +\tau_{1}\Lambda_{0}^{(\theta)}=0,
\eeqa
where $\Lambda_{0}^{(\theta)}=x_1 \Lambda_{10}^{(\theta)}+x_2\Lambda_{20}^{(\theta)}$ and $\Lambda_{1}^{(\theta)}=x_1 \Lambda_{11}^{(\theta)}+x_2\Lambda_{21}^{(\theta)}$. Here, we have introduced the quantities
\beq
\label{a5}
\Lambda_{10}^{(\theta)}=\gamma_1^*\theta^{-1}\tau_1-3\gamma_1^*\theta^{-2}, \quad \Lambda_{20}^{(\theta)}=\gamma_2^*\theta^{-1}\tau_2-3\gamma_2^*\theta^{-2},
\eeq
\beq
\label{a6}
\Lambda_{11}^{(\theta)}=-2\gamma_1^*-\frac{3}{2}\zeta_{10}^*+\tau_1^{-1/2}\frac{M_1^{1/2}}{x_2M_2}
\left(\frac{\partial\zeta'_1}{\partial\beta}\right)_{x_1,\phi}
\eeq
\beq
\label{a7}
\Lambda_{21}^{(\theta)}=2\frac{x_1}{x_2}\gamma_2^*+\frac{3}{2}\frac{x_1}{x_2}\zeta_{20}^*+
\frac{M_1}{x_2M_2^{3/2}}\frac{\tau_2^{3/2}}{\tau_1^{2}}\left(\frac{\partial\zeta_2'}{\partial\beta}\right)_{x_1,\phi}.
\eeq
In Eqs.\ \eqref{a4}--\eqref{a7}, although the subscript $s$ has been omitted for the sake of simplicity, it is understood that all the quantities are evaluated in the steady state. As for dilute driven granular mixtures \cite{KG13}, an analysis of the solutions to Eq. \eqref{a4} shows that in general one of the roots leads to unphysical behavior of the diffusion coefficients. We take the other root as
the physical root of the quadratic equation \eqref{a4}.

Once the derivative $\Delta_{\theta,1}$ is known, we can determine the remaining derivatives in a similar way. In order to get $(\partial\tau_1/\partial\lambda_1)_{\theta,x_1,\phi}$, we take first the derivative of Eq.\ \eqref{a1} with respect to $\lambda_1$ and then consider the steady-state conditions. The final result is
\beq
\label{a8}
\Delta_{\lambda_1,1}=-\frac{\tau_{1}\Lambda_{0}^{(\lambda_1)}-
\Lambda_{10}^{(\lambda_1)}+\theta \Lambda_{0}^{(\lambda_1)}\Delta_{\theta,1}}{\theta
\Lambda_{1}^{(\theta)}\Delta_{\theta,1}+\tau_{1}\Lambda_{1}^{(\theta)}
-\Lambda_{11}^{(\theta)}},
\eeq
where $\Lambda_{0}^{(\lambda)}=x_{1}\Lambda_{10}^{(\lambda_1)}+x_{2}\Lambda_{20}^{(\lambda_1)}$, and
\beq
\label{a9}
\Lambda_{10}^{(\lambda_1)}=2\theta^{-1/2}\left(\theta^{-1}-\tau_1\right),\quad \Lambda_{20}^{(\lambda)}=2\frac{R_2}{R_1}\theta^{-1/2}\left(\theta^{-1}-\tau_2\right).
\eeq

Analogously, the derivative $(\partial\tau_1/\partial x_1)_{\theta,\lambda_1,\phi}$ in the steady state is
\beq
\label{a10}
\Delta_{x_1,1}=-\frac{\tau_{1}\Lambda_{0}^{(x_1)}-\Lambda_{10}^{(x_1)}+\theta
\Lambda_{0}^{(x_1)}\Delta_{\theta,1}}{\theta
\Lambda_{1}^{(\theta)}\Delta_{\theta,1}+\tau_{1}\Lambda_{1}^{(\theta)}
-\Lambda_{11}^{(\theta)}},
\eeq
where $\Lambda_{0}^{(x_1)}=x_{1}\Lambda_{10}^{(x_1)}+x_{2}\Lambda_{20}^{(x_1)}$, and
\beq
\label{a11}
\Lambda_{10}^{(x_1)}=-\tau_1^{3/2}M_1^{-1/2}\bigg(\frac{\partial\zeta'_1}{\partial x_1}\bigg)_{\beta,\phi},
\eeq
\beqa
\label{a12}
\Lambda_{20}^{(x_1)}&=& 2\frac{\gamma_1^*}{x_2}\left(\theta^{-1}-\tau_1\right)-2\frac{\gamma_2^*}{x_2}\left(\theta^{-1}-\tau_2\right)\nonumber\\
& &  -\frac{1-\tau_1}{x_2^2}\gamma_2^*+\frac{3}{2}\frac{\tau_1-\frac{\tau_2}{3}}{x_2}\zeta_{2,0}^*-\frac{\tau_1}{x_2}\zeta_{1,0}^*\nonumber\\
& & -\tau_2^{3/2}M_2^{-1/2}\bigg(\frac{\partial\zeta'_2}{\partial x_1}\bigg)_{\beta,\phi}.
\eeqa

Finally, in the steady state, the derivative $(\partial \tau_1/\partial \phi)_{\theta,x_1,\lambda_1}$ is
\beq
\label{a13}
\Delta_{\phi,1}=-\frac{\tau_{1}\Lambda_{0}^{(\phi)}-\Lambda_{10}^{(\phi)}+\theta
\Lambda_{0}^{(\phi)}\Delta_{\theta,1}}{\theta
\Lambda_{1}^{(\theta)}\Delta_{\theta,1}+\tau_{1}\Lambda_{1}^{(\theta)}
-\Lambda_{11}^{(\theta)}},
\eeq
where $\Lambda_{0}^{(\phi)}=x_{1}\Lambda_{10}^{(\phi)}+x_{2}\Lambda_{20}^{(\phi)}$, and
\beq
\label{a14}
\Lambda_{10}^{(\phi)}=-\tau_1^{3/2}M_1^{-1/2}\bigg(\frac{\partial\zeta_1'}{\partial \phi}\bigg)_{x_1,\beta},
\eeq
\beq
\label{a15}
\Lambda_{20}^{(\phi)}=-\tau_2^{3/2}M_2^{-1/2}\bigg(\frac{\partial\zeta_2'}{\partial \phi}\bigg)_{x_1,\beta}.
\eeq

\section{Some technical details on the first-order Chapman--Enskog solution}
\label{appB}

\begin{widetext}
To first order in the spatial gradients, the distribution function $f_i^{(1)}$ obeys the Enskog kinetic equation
\beq
\label{b1}
\partial_t^{(0)}f_i^{(1)}-\gamma_i\frac{\partial}{\partial\mathbf{v}}\cdot\mathbf{V}f_i^{(1)}-\frac{\gamma_i T_{\text{ex}}}{m_i}\frac{\partial^2 f_i^{(1)}}{\partial v^2}=-\left(D_t^{(1)}+\mathbf{V}\cdot\nabla\right)f_i^{(0)}+\gamma_i\Delta\mathbf{U}\cdot\frac{\partial f_i^{(0)} }{\partial\mathbf{v}}-\mathbf{g}\cdot \frac{\partial f_i^{(0)}}{\partial\mathbf{v}}+\sum_{j=1}^2\; J_{ij}^{(1)}[f_i,f_j],
\eeq
where $D_t^{(1)}\equiv \partial_t^{(1)}+\mathbf{U}\cdot \nabla$ and $J_{ij}^{(1)}[f_i,f_j]$ denotes the first-order contribution to the expansion of the Enskog collision operator in spatial gradients. To obtain $J_{ij}^{(1)}[f_i,f_j]$ one needs the expansions \cite{GDH07,G19}
\begin{equation}
\label{b2}
\chi_{ij}\left(\mathbf{r},\mathbf{r}\pm\boldsymbol{\sigma}_{ij}|\left\{n_\ell\right\}\right)\rightarrow\sum_{\ell=1}^2
\chi_{ij}^{(0)}\left(1\pm\frac{1}{2}\left(n_{\ell}\frac{\partial \ln \chi_{ij}^{(0)}}{\partial n_{\ell}}+I_{ij\ell}\right)\boldsymbol{\sigma}_{ij}\cdot\nabla\ln n_{\ell}\right),
\end{equation}
\begin{equation}
\label{b3}
f_j^{(0)}(\mathbf{r}\pm\boldsymbol{\sigma}_{ij})\rightarrow \sum_{\ell=1}^2n_{\ell}\frac{\partial f_j^{(0)}}{\partial n_{\ell}}\boldsymbol{\sigma}_{ij}\cdot\nabla\ln n_{\ell} -\frac{\partial f_j^{(0)}}{\partial V_\beta}(\boldsymbol{\sigma}_{ij}\cdot\nabla)U_\beta+T\frac{\partial f_j^{(0)}}{\partial T}\boldsymbol{\sigma}_{ij}\cdot\nabla\ln T.
\end{equation}
In Eq.\ \eqref{b3}, the quantities $I_{ij\ell}$ are defined in terms of the functional derivative of the (local) pair distribution function $\chi_{ij}$ with respect to the (local) partial densities $n_\ell$. These quantities are the origin of the primary difference between the so-called standard and revised Enskog kinetic theories for ordinary mixtures \cite{BE73c,LCK83}. The explicit forms of $I_{ij\ell}$ for a binary mixture of hard disks ($d=2$) or spheres ($d=3$) have been provided in the Appendix A of Ref.\ \cite{G11}. Taking into account the expansions \eqref{b2} and \eqref{b3}, the operator $J_{ij}^{(1)}[f_i,f_j]$ can be written as
\beqa
\label{b4}
\sum_{j=1}^2J_{ij}^{(1)}[f_i,f_j]&\rightarrow& -\sum_{j=1}^2\sum_{\ell=1}^2\left\{\boldsymbol{\mathcal{K}}_{ij}\left[n_{\ell}\frac{\partial f_j^{(0)}}{\partial n_{\ell}}\right]+\frac{1}{2}\left(n_{\ell}\frac{\partial \ln \chi_{ij}^{(0)}}{\partial n_{\ell}}+I_{ij\ell}\right)\boldsymbol{\mathcal{K}}_{ij}\left[f_j^{(0)}\right]\right\}\cdot\nabla\ln n_\ell\nonumber\\
& &-\sum_{j=1}^2 \boldsymbol{\mathcal{K}}_{ij}\left[T\frac{\partial f_j^{(0)}}{\partial T}\right]\cdot\nabla\ln T+\frac{1}{2}\sum_{j=1}^2\mathcal{K}_{ij,\lambda}\left[\frac{\partial f_j^{(0)}}{\partial V_\beta}\right]\left(\partial_\lambda U_\beta+\partial_\beta U_\lambda-\frac{2}{d}\delta_{\lambda\beta}\nabla\cdot\mathbf{U}\right)\nonumber\\
& &+\frac{1}{d}\sum_{j=1}^2\mathcal{K}_{ij,\lambda}\left[\frac{\partial f_j^{(0)}}{\partial V_\lambda}\right]\nabla\cdot\mathbf{U}
+\sum_{j=1}^2\left(J_{ij}^{(0)}\left[f_i^{(1)},f_j^{(0)}\right]+J_{ij}^{(0)}\left[f_i^{(0)},f_j^{(1)}\right]\right),
\eeqa
where the operator $\boldsymbol{\mathcal{K}}_{ij}[X_j]$ is given by \cite{GDH07,G19}
\beq
\label{b5}
\boldsymbol{\mathcal{K}}_{ij}[X_j]=\sigma_{ij}^d\chi_{ij}^{(0)}\int\dd\mathbf{v}_2\int\dd\widehat{\boldsymbol{\sigma}}
\Theta(\widehat{\boldsymbol{\sigma}}\cdot\mathbf{g}_{12})\left(\widehat{\boldsymbol{\sigma}}\cdot\mathbf{g}_{12}\right)
\widehat{\boldsymbol{\sigma}}\left[\alpha_{ij}^{-2}f_i^{(0)}(\mathbf{v}_1'')X_j(\mathbf{v}_2'')+f_i^{(0)}
(\mathbf{v}_1)X_j(\mathbf{v}_2)\right].
\eeq
As for monocomponent granular suspensions \cite{GGG19a}, upon deriving  Eq.\ \eqref{b4} use has been made of the symmetry property $\mathcal{K}_{ij,\lambda}[\partial_{V_\beta}f_j^{(0)}]=\mathcal{K}_{ij,\beta}[\partial_{V_\lambda}f_j^{(0)}]$ that follows from the isotropy in velocity space of the zeroth-order distributions $f_i^{(0)}$.

To first-order, the balance equations are
\beq
\label{b6}
D_t^{(1)}n_i=-n_i\nabla\cdot\mathbf{U},\quad D_t^{(1)}\mathbf{U}=-\rho^{-1}\nabla p-\Delta\mathbf{U}\sum_{i=1}^2\frac{\rho_i}{\rho}\gamma_i+\mathbf{g}+\rho^{-1}\left(\gamma_1-\gamma_2\right)\mathbf{j}_1^{(1)},
\eeq
\beq
\label{b7}
D_t^{(1)}T=-\frac{2p}{dn}\nabla\cdot\mathbf{U}-\zeta^{(1)}T-2\sum_{i=1}^2\gamma_i x_iT_i^{(1)}.
\eeq
Here, $p$ is given by Eq.\ \eqref{4.8} and $\zeta^{(1)}$ is the first order contribution to the cooling rate. Since $\zeta^{(1)}$ is a scalar, corrections to first-order in the gradients can arise only from $\nabla \cdot \mathbf{U}$ since $\nabla n_i$ and $\nabla T$ are vectors and the tensor $\partial_\lambda U_\beta+\partial_\beta U_\lambda-\frac{2}{d}\delta_{\lambda\beta}\nabla\cdot\mathbf{U}$ is a traceless tensor. Thus, $\zeta^{(1)}=\zeta_U \nabla \cdot \mathbf{U}$, where $\zeta_U$ can be decomposed as $\zeta_U=\zeta^{(1,0)}+\zeta^{(1,1)}$. The coefficient $\zeta^{(1,0)}$ can be evaluated explicitly with the result \cite{GHD07}
\beq
\label{b8}
\zeta^{(1,0)}=-\frac{3}{n T}\frac{\pi^{d/2}}{d^2\Gamma\left(\frac{d}{2}\right)}\sum_{i=1}^2\sum_{j=1}^2 n_i n_j \mu_{ji}\sigma_{ij}^d \chi_{ij}^{(0)}T_i^{(0)}(1-\al_{ij}^2).
\eeq
On the other hand, the coefficient $\zeta^{(1,1)}$ is given in terms of the first-order distributions $f_i^{(1)}$. Its expression will be displayed latter. In addition,  according to Eq.\ \eqref{4.8}, $\nabla p$ can be written as
\beq
\label{b9}
\nabla p=\sum_{i=1}^2 n_i \frac{\partial p}{\partial n_i} \nabla \ln n_i+p\left(1+\theta \frac{\partial \ln p^*}{\partial \theta}\right)\nabla \ln T,
\eeq
where we recall that $p^*=p/nT$.

The right-hand side of Eq.\ \eqref{b1} can be evaluated by using Eqs.\ \eqref{b6}--\eqref{b9} and the expansion \eqref{b4} of the Enskog operator. With these results, the corresponding kinetic equation for $f_i^{(1)}$ reads
\beqa
\label{b10}
& &\partial_t^{(0)}f_i^{(1)}-\gamma_i\frac{\partial}{\partial\mathbf{v}}\cdot\mathbf{V}f_i^{(1)}-\frac{\gamma_i T_{\text{ex}}}{m_i}\frac{\partial^2 f_i^{(1)}}{\partial v^2}-\left(T \zeta^{(1,1)}+2\sum_{j=1}^2 x_j \gamma_j \varpi_j\right)\frac{\partial f_i^{(0)}}{\partial T}\nabla\cdot\mathbf{U}-\rho^{-1}\left(\gamma_1-\gamma_2\right)\mathbf{j}_1^{(1)}\cdot \frac{\partial f_i^{(0)}}{\partial \mathbf{V}}\nonumber\\
& &
-\sum_{j=1}^2\left(J_{ij}^{(0)}\left[f_i^{(1)},f_j^{(0)}\right]+J_{ij}^{(0)}\left[f_i^{(0)},f_j^{(1)}\right]\right)
= \mathbf{A}_i\cdot\nabla\ln T
+\sum_{j=1}^2\mathbf{B}_{ij}\cdot\nabla\ln n_j+C_{i,\lambda\beta}\frac{1}{2}\left(\partial_\lambda U_\beta+\partial_\beta U_\lambda-\frac{2}{d}\delta_{\lambda\beta}\nabla\cdot\mathbf{U}\right)\nonumber\\
& & +D_i\nabla\cdot\mathbf{U}+\mathbf{E}_i \cdot \Delta \mathbf{U},
\eeqa
where
\beq
\label{b11}
\mathbf{A}_i(\mathbf{V})=-\mathbf{V}T\frac{\partial f_i^{(0)}}{\partial T}-\frac{p}{\rho}\left(1+\theta \frac{\partial \ln p^*}{\partial \theta}\right)\frac{\partial f_i^{(0)}}{\partial\mathbf{V}}-\sum_{j=1}^2\boldsymbol{\mathcal{K}}_{ij}\left[T\frac{\partial f_j^{(0)}}{\partial T}\right],
\eeq
\beq
\label{b12}
\mathbf{B}_{ij}(\mathbf{V})=-\mathbf{V}n_j\frac{\partial f_i^{(0)}}{\partial n_j}-\frac{n_j}{\rho}\frac{\partial p}{\partial n_j}\frac{\partial f^{(0)}_i}{\partial\mathbf{V}}-\sum_{\ell=1}^2\left\{\boldsymbol{\mathcal{K}}_{i\ell}\left[n_{j}\frac{\partial f_\ell^{(0)}}{\partial n_{j}}\right]+\frac{1}{2}\left(n_{j}\frac{\partial \ln \chi_{i\ell}^{(0)}}{\partial n_{j}}+I_{i\ell j}\right)\boldsymbol{\mathcal{K}}_{i\ell}\left[f_\ell^{(0)}\right]\right\},
\eeq
\beq
\label{b13}
C_{i,\beta\lambda}(\mathbf{V})=V_\lambda\frac{\partial f_i^{(0)}}{\partial V_\beta}+\sum_{j=1}^2\mathcal{K}_{ij,\lambda}\left[\frac{\partial f_j^{(0)}}{\partial V_\beta}\right],
\eeq
\beq
\label{b14}
D_i(\mathbf{V})=\frac{1}{d}\frac{\partial}{\partial\mathbf{V}}\cdot\left(\mathbf{V}f_i^{(0)}\right)+\left(\zeta^{(1,0)}
+\frac{2}{d}p^*\right)T\frac{\partial f_i^{(0)}}{\partial T}-f_i^{(0)}+\sum_{j=1}^2\left\{n_j\frac{\partial f^{(0)}_i}{\partial n_j}+\frac{1}{d}\mathcal{K}_{ij,\lambda}\left[\frac{\partial f_j^{(0)}}{\partial V_\lambda}\right]\right\},
\eeq
\beq
\label{b14a}
\mathbf{E}_i\left(\mathbf{V}\right)=\left(\gamma_i-\sum_{j=1}^2 \frac{\rho_j}{\rho}\gamma_j\right)\frac{\partial f^{(0)}_i}{\partial\mathbf{V}}.
\eeq
Note that in Eq.\ \eqref{b10}, $\zeta_1^{(1)}$ and $\varpi_i$ are functionals of the first-order distributions $f_i^{(1)}$. In Eq.\ \eqref{b12}, the derivative $\partial f_i^{(0)}/\partial n_j$ can be more explicitly written when one takes into account the scaling solution \eqref{3.12}:
\beq
\label{b14.1}
n_j\frac{\partial f_i^{(0)}}{\partial n_j}=\delta_{ij}f_j^{(0)}+n_jf_i^{(0)}\left(\frac{\partial \ln \varphi_i}{\partial x_1}\frac{\partial x_1}{\partial n_j}+\frac{\partial \ln \varphi_i}{\partial \lambda_1}\frac{\partial \lambda_1}{\partial n_j}+\frac{\partial \ln \varphi_i}{\partial \phi}\frac{\partial \phi}{\partial n_j}\right),
\eeq
where
\beq
\label{b14.2}
n_j\frac{\partial x_1}{\partial n_j}=x_j\left(x_2 \delta_{1j}-x_1 \delta_{2j}\right), \quad n_j\frac{\partial \phi}{\partial n_j}=\phi_j,
\eeq
\beq
\label{b14.3}
n_j\frac{\partial \lambda_1}{\partial n_j}=\lambda_1\left(\phi_j\frac{\partial \ln R_1}{\partial \phi}+\frac{\partial \ln R_1}{\partial x_1}n_j\frac{\partial x_1}{\partial n_j}\right)-\lambda_1 x_j-\lambda_1\frac{\rho_j}{\rho}.
\eeq

The solution to Eq.\ \eqref{b10} is given by Eq.\ \eqref{4.15}. Because of the gradients $\nabla n_i$, $\nabla T$, and $\nabla\cdot \mathbf{U}$ as well as the traceless tensor $\partial_\lambda U_\beta+\partial_\beta U_\lambda-\frac{2}{d}\delta_{\lambda\beta}\nabla\cdot\mathbf{U}$ are all independents, substitution of the form \eqref{4.15} into Eq.\ \eqref{b10} leads to the following set of linear integral equations for the unknowns $\boldsymbol{\mathcal{A}}_i(\mathbf{V})$, $\boldsymbol{\mathcal{B}}_{ij}(\mathbf{V})$, $\mathcal{C}_{i,\lambda\beta}(\mathbf{V})$, and $\mathcal{D}_i(\mathbf{V})$:
\beqa
\label{n1}
& &\Lambda^{(0)} T\partial_T\boldsymbol{\mathcal{A}}_i-\left[2\sum_{j=1}^2 \gamma_j x_j\left(\theta^{-1}+\theta\frac{\partial\tau_j}{\partial\theta}\right)+\frac{1}{2}\zeta^{(0)}+\zeta^{(0)} \theta
\frac{\partial \ln \zeta_0^*}{\partial \theta}\right]\boldsymbol{\mathcal{A}}_i-\gamma_i\frac{\partial}{\partial\mathbf{v}}\cdot\mathbf{V}\boldsymbol{\mathcal{A}}_i
-\frac{\gamma_i T_{\text{ex}}}{m_i}\frac{\partial^2}{\partial v^2}\boldsymbol{\mathcal{A}}_i\nonumber\\
& &+ \left(\gamma_2-\gamma_1\right) D_1^T\frac{\partial f_i^{(0)}}{\partial\mathbf{V}}-\sum_{j=1}^2\left(J_{ij}^{(0)}\left[\boldsymbol{\mathcal{A}}_i,f_j^{(0)}\right]
+J_{ij}^{(0)}\left[f_i^{(0)},\boldsymbol{\mathcal{A}}_j\right]\right)=\mathbf{A}_i,
\eeqa
\beqa
\label{n2}
& & \Lambda^{(0)} T\partial_T\boldsymbol{\mathcal{B}}_{ij}-\gamma_i\frac{\partial}{\partial\mathbf{v}}\cdot\mathbf{V}\boldsymbol{\mathcal{B}}_{ij}
-\frac{\gamma_i T_{\text{ex}}}{m_i}\frac{\partial^2}{\partial v^2}\boldsymbol{\mathcal{B}}_{ij}+\left(\gamma_2-\gamma_1\right)\frac{m_1\rho_j}{\rho^2}D_{1j}\frac{\partial f_i^{(0)}}{\partial\mathbf{V}}-\sum_{\ell=1}^2\left(J_{i\ell}^{(0)}\left[\boldsymbol{\mathcal{B}}_{ij},
f_\ell^{(0)}\right]+J_{i\ell}^{(0)}\left[f_i^{(0)},\boldsymbol{\mathcal{B}}_{\ell j}\right]\right)\nonumber\\
& & =\mathbf{B}_{ij}+\left[n_j\frac{\partial\zeta^{(0)}}{\partial n_j}-2n_j\sum_{\ell=1}^2\left\{\gamma_\ell x_\ell\left[ \left(\theta^{-1}-\tau_\ell\right)\left(\frac{\partial \ln \gamma_\ell}{\partial n_j}+\frac{\partial \ln x_\ell}{\partial n_j}\right)-\left(\frac{\partial\tau_\ell}{\partial x_1}\frac{\partial x_1}{\partial n_j}+\frac{\partial\tau_\ell}{\partial\lambda_1}\frac{\partial\lambda_1}{\partial n_j}+\frac{\partial\tau_\ell}{\partial\phi}\frac{\partial\phi}{\partial n_j}\right)\right]\right\}\right]\boldsymbol{\mathcal{A}}_i,
\nonumber\\
\eeqa
\beq
\label{n3}
\Lambda^{(0)} T\partial_T\mathcal{C}_{i,\lambda\beta}-\gamma_i\frac{\partial}{\partial\mathbf{v}}\cdot\mathbf{V}\mathcal{C}_{i,\lambda\beta}
-\frac{\gamma_i T_{\text{ex}}}{m_i}\frac{\partial^2}{\partial v^2}\mathcal{C}_{i,\lambda\beta}-\sum_{j=1}^2\left(J_{ij}^{(0)}\left[\mathcal{C}_{i,\lambda\beta},f_j^{(0)}\right]
+J_{ij}^{(0)}\left[f_i^{(0)},\mathcal{C}_{j,\lambda\beta}\right]\right)=C_{i,\lambda\beta},
\eeq
\beq
\label{n4}
\Lambda^{(0)} T\partial_T\mathcal{D}_{i}-\gamma_i\frac{\partial}{\partial\mathbf{v}}\cdot\mathbf{V}\mathcal{D}_i-\frac{\gamma_i T_{\text{ex}}}{m_i}\frac{\partial^2}{\partial v^2}\mathcal{D}_i-\left(T \zeta^{(1,1)}+2\sum_{j=1}^2 \gamma_j x_j \varpi_j\right)\frac{\partial f_i^{(0)}}{\partial T}-\sum_{j=1}^2\left(J_{ij}^{(0)}\left[\mathcal{D}_i,f_j^{(0)}\right]+J_{ij}^{(0)}
\left[f_i^{(0)},\mathcal{D}_j\right]\right)=D_i,
\eeq
\beq
\label{n5}
\Lambda^{(0)} T\partial_T\boldsymbol{\mathcal{E}}_{i}-\gamma_i\frac{\partial}{\partial\mathbf{v}}\cdot\mathbf{V}\boldsymbol{\mathcal{E}}_i-\frac{\gamma_i T_{\text{ex}}}{m_i}\frac{\partial^2}{\partial v^2}\boldsymbol{\mathcal{E}}_i+\rho^{-1}\left(\gamma_2-\gamma_1\right)D^U_1\frac{\partial f_i^{(0)}}{\partial \mathbf{V}}-\sum_{j=1}^2\left(J_{ij}^{(0)}\left[\boldsymbol{\mathcal{E}}_i,f_j^{(0)}\right]+J_{ij}^{(0)}
\left[f_i^{(0)},\boldsymbol{\mathcal{E}}_j\right]\right)=\mathbf{E}_i,
\eeq
where $\Lambda^{(0)}$ is defined by Eq.\ \eqref{4.10}. Upon deriving the above integral equations use has been made of the constitutive equation \eqref{5.1} for the mass flux $\mathbf{j}_1^{(1)}$ and the result
\beqa
\label{b15}
& & \partial_t^{(0)}\nabla\ln T=\nabla\partial_t^{(0)}\ln T=\nabla\bigg(2\sum_{i=1}^2\gamma_i x_i\left(\theta^{-1}-\tau_i\right)-\zeta^{(0)}\bigg)\nonumber\\
&=& -\sum_{j=1}^2\left[n_j\frac{\partial\zeta^{(0)}}{\partial n_j}-2n_j\sum_{\ell=1}^2\left\{\gamma_\ell x_\ell\left[ \left(\theta^{-1}-\tau_\ell\right)\left(\frac{\partial \ln \gamma_\ell}{\partial n_j}+\frac{\partial \ln x_\ell}{\partial n_j}\right)-\left(\frac{\partial\tau_\ell}{\partial x_1}\frac{\partial x_1}{\partial n_j}+\frac{\partial\tau_\ell}{\partial\lambda_1}\frac{\partial\lambda_1}{\partial n_j}+\frac{\partial\tau_\ell}{\partial\phi}\frac{\partial\phi}{\partial n_j}\right)\right]\right\}\right]\nabla\ln n_j\nonumber\\
& &-\left[2\sum_{i=1}^2 \gamma_i x_i\left(\theta^{-1}+\theta\frac{\partial\tau_i}{\partial\theta}\right)+\frac{1}{2}\zeta^{(0)}+\zeta^{(0)} \theta
\frac{\partial \ln \zeta_0^*}{\partial \theta}\right]\nabla\ln T.
\eeqa
Moreover, since $\zeta^{(1,1)}$ is coupled to $\mathcal{D}_i$, its explicit form can be easily identified after expanding the expression \eqref{2.25} of the cooling rate to first order. The result is \cite{GHD07}
\beq
\label{b16}
\zeta^{(1,1)}=\frac{1}{nT}\frac{\pi^{(d-1)/2}}{d\Gamma\left(\frac{d+3}{2}\right)}\sum_{i=1}^2\sum_{j=1}^2 \sigma_{ij}^{d-1} \chi_{ij}^{(0)}m_{ij} (1-\al_{ij}^2)\int\dd \mathbf{v}_1\int\dd \mathbf{v}_2 \; g_{12}^3\; f_i^{(0)}(\mathbf{V}_1)\mathcal{D}_j (\mathbf{V}_2).
\eeq
The integral equations \eqref{4.16}--\eqref{4.19} can be obtained from Eqs.\ \eqref{n1}--\eqref{n4} when the steady state condition ($\Lambda^{(0)}=0$) is assumed.


\section{Algebraic equations defining the transport coefficients}
\label{appC}

In this Appendix, we display the set of algebraic equations defining the diffusion transport coefficients, the shear viscosity coefficient, and the first-order contributions to the partial temperatures. In the case of the diffusion coefficients $D_i^T$, $D_{ij}$, and $D_i^U$, the set of algebraic equations are, respectively, given by
\beqa
\label{5.19}
& & \sum_{j=1}^2\left\{\nu_{ij}+\left(\gamma_2-\gamma_1\right)\frac{\rho_i}{\rho}\delta_{1j}-\left[2\sum_{\ell=1}^2\gamma_\ell x_\ell\left(\theta^{-1}-\theta \Delta_{\theta,\ell}\right)+\frac{1}{2}\zeta^{(0)}+\zeta^{(0)}\theta \frac{\partial\ln \zeta_0^*}{\partial \theta}-\gamma_i\right]\delta_{ij}\right\}D_j^T \nonumber\\
&=&-\frac{p\rho_i}{\rho^2}\left(1+\theta \frac{\partial \ln p^*}{\partial \theta}-\frac{\rho n_iT_i^{(0)}}{p\rho_i}\right)
+T\frac{n_i}{\rho}\theta\Delta_{\theta,i}
+\frac{\pi^{d/2}}{d\Gamma\left(\frac{d}{2}\right)}\frac{n_i T}{\rho}\sum_{j=1}^2n_j
\mu_{ij}\chi_{ij}^{(0)}\sigma_{ij}^d \left(1+\alpha_{ij}\right)\nonumber\\
& & \times\left(\tau_j+\theta\Delta_{\theta,j}\right),
\eeqa
\beqa
\label{5.20}
& & \sum_{\ell=1}^2\left[\nu_{i\ell}+\left(\gamma_2-\gamma_1\right)\frac{\rho_i}{\rho}\delta_{1\ell}+\gamma_i\delta_{i\ell}\right]m_{\ell} D_{\ell j}= \frac{\rho T}{\rho_j}\Bigg\{n_j \tau_j\delta_{ij}+n\Bigg[n_j \frac{\partial x_1}{\partial n_j}x_i\Delta_{x_1,i}
+n_j \frac{\partial \lambda_1}{\partial n_j}x_i\Delta_{\lambda_1,i}+x_i\phi_j \Delta_{\phi,i}\Bigg]\Bigg\}\nonumber\\
&+&\frac{\rho^2}{m_j}
\left[\frac{\partial\zeta^{(0)}}{\partial n_j}-2\sum_{\ell=1}^2\left\{\gamma_\ell x_\ell\left[ \left(\theta^{-1}-\tau_\ell\right)\left(\frac{\partial \ln \gamma_\ell}{\partial n_j}+\frac{\partial \ln x_\ell}{\partial n_j}\right)-\left(\frac{\partial x_1}{\partial n_j}\Delta_{x_1,\ell}+\frac{\partial\lambda_1}{\partial n_j}\Delta_{\lambda_1,\ell}+\frac{\phi_j}{n_j}\Delta_{\phi,\ell}\right)\right]\right\}\right]D_i^T\nonumber\\
& & -\frac{\rho_i}{m_j}\frac{\partial p}{\partial n_j}+\frac{\rho T}{\rho_j}\frac{\pi^{d/2}}{d\Gamma\left(\frac{d}{2}\right)}\sum_{\ell=1}^2 n_i n_\ell \sigma_{i \ell}^d\chi_{i\ell}^{(0)} m_{i\ell}
(1+\al_{i\ell}\Bigg\{\Bigg[\delta_{j\ell}+\frac{1}{2}\left(n_{j}\frac{\partial \ln \chi_{i\ell}^{(0)}}{\partial n_{j}}+I_{i\ell j}\right)\Bigg]
\Bigg(\frac{\tau_i}{m_i}+\frac{\tau_\ell}{m_\ell}\Bigg)
\nonumber\\
& & +\frac{n_j}{m_\ell}\frac{\partial x_1}{\partial n_j}\Delta_{x_1,\ell}+\frac{n_j}{m_\ell}\frac{\partial \lambda_1}{\partial n_j}\Delta_{\lambda_1,\ell}+\frac{\phi_j}{m_\ell}\Delta_{\phi,\ell}\Bigg\},
\eeqa
\beq
\label{5.21}
\sum_{j=1}^2\left[\nu_{ij}+\left(\gamma_2-\gamma_1\right)\frac{\rho_i}{\rho}\delta_{1j}+\gamma_i\delta_{ij}\right] D^U_{j}= \rho_i\left(\gamma_i-\sum_{j=1}^2\frac{\rho_j}{\rho}\gamma_j\right).
\eeq
Here, the derivatives $\partial x_1/\partial n_j$ and $\partial \lambda_1/\partial n_j$ are given by Eqs.\ \eqref{b14.2} and \eqref{b14.3}, respectively, and the collision frequencies $\nu_{ij}$ appearing in Eqs.\ \eqref{5.19}--\eqref{5.21} are defined as
\beq
\label{5.21a}
\nu_{ii}=-\frac{m_i}{dn_iT_i}\int\dd\mathbf{v}\; \mathbf{V}\cdot J_{ij}^{(0)}[f_{i,\text{M}}\mathbf{V},f_j^{(0)}],
\eeq
\beq
\label{5.21b}
\nu_{ij}=-\frac{m_i}{dn_jT_j}\int\dd\mathbf{v}\; \mathbf{V}\cdot J_{ij}^{(0)}[f_i^{(0)},f_{j,\text{M}}\mathbf{V}],
\eeq
for $i \neq j$.  Note that the self-collision terms of $\nu_{ii}$ arising from $J_{ii}^{(0)}[f_{i\text{M}}\mathbf{V},f_i^{(0)}]$ do not occur in Eq.\ \eqref{5.21a} since they conserve momentum for the component $i$. In addition, upon deriving Eqs.\ \eqref{5.19} and \eqref{5.20}, use has been made of the results
\beq
\label{5.21.1}
\int \dd{\bf v} m_i \mathbf{V}\cdot \boldsymbol{\mathcal {K}}_{ij}\Bigg[T\frac{\partial f_j^{(0)}}{\partial T}\Bigg]=
\frac{\pi^{d/2}}{\Gamma\left(\frac{d}{2}\right)}n_i n_j \sigma_{ij}^d \chi_{ij}^{(0)}\mu_{ij}(1+\al_{ij})
T_j^{(0)}\left(1+\frac{\theta}{\tau_j}\Delta_{\theta,j}\right),
\eeq
\beqa
\label{5.21.2}
& & \int \dd{\bf v} m_i \mathbf{V}\cdot \left\{\boldsymbol{\mathcal{K}}_{i\ell}\left[n_{j}\frac{\partial f_\ell^{(0)}}{\partial n_{j}}\right]+\frac{1}{2}\left(n_{j}\frac{\partial \ln \chi_{i\ell}^{(0)}}{\partial n_{j}}+I_{i\ell j}\right)\boldsymbol{\mathcal{K}}_{i\ell}\left[f_\ell^{(0)}\right]\right\}=\frac{\pi^{d/2}}{\Gamma\left(\frac{d}{2}\right)}n_i T n_\ell \sigma_{i\ell}^d \chi_{i\ell}^{(0)}m_{i\ell}(1+\al_{i\ell})
\nonumber\\
& &\times \Bigg\{\Bigg[\delta_{j\ell}
+\frac{1}{2}\left(n_{j}\frac{\partial \ln \chi_{i\ell}^{(0)}}{\partial n_{j}}+I_{i\ell j}\right) \Bigg]
\Bigg(\frac{\tau_i}{m_i}+\frac{\tau_\ell}{m_\ell}\Bigg)+\frac{n_j}{m_\ell}\frac{\partial x_1}{\partial n_j}\Delta_{x_1,\ell}+\frac{n_j}{m_\ell}\frac{\partial \lambda_1}{\partial n_j}
\Delta_{\lambda_1,\ell}+\frac{\phi_j}{m_\ell}\Delta_{\phi,\ell}\Bigg\},
\eeqa
where $f_i^{(0)}$ has been replaced by $f_{i,\text{M}}$. The explicit forms of the collision frequencies $\nu_{ii}$ and $\nu_{ij}$ can be also easily obtained by considering the latter replacement. They are given by \cite{GM07}
\beq
\label{5.21c}
\nu_{ii}=\frac{2\pi^{(d-1)/2}}{d\Gamma\left(\frac{d}{2}\right)}n_j\sigma_{ij}^{d-1}\chi_{ij}^{(0)}
\mu_{ji}v_0\left(1+\alpha_{ij}\right)\left(\frac{\beta_i+\beta_j}{\beta_i\beta_j}\right)^{1/2},
\eeq
\beq
\label{5.21d}
\nu_{ij}=-\frac{2\pi^{(d-1)/2}}{d\Gamma\left(\frac{d}{2}\right)}n_i\sigma_{ij}^{d-1}\chi_{ij}^{(0)}\mu_{ij}
v_0\left(1+\alpha_{ij}\right)\left(\frac{\beta_i+\beta_j}{\beta_i\beta_j}\right)^{1/2}.
\eeq
We recall that $i\neq j$ in Eqs.\ \eqref{5.21c} and \eqref{5.21d}. With these results, the explicit form of $D_1^T$ can be written as
\beqa
\label{5.21.4}
D_1^T&=&\Bigg[\nu_D+\frac{\rho_1\gamma_2+\rho_2\gamma_1}{\rho}-2\sum_{j=1}^2x_j\gamma_j \left(\theta^{-1}-\theta \Delta_{\theta,j}\right)-\zeta^{(0)}\left(\frac{1}{2}+\theta \frac{\partial\ln \zeta_0^*}{\partial \theta}\right)\Bigg]^{-1}
\Bigg\{\frac{T n_1}{\rho}\theta \Delta_{\theta,1}\nonumber\\
& & -\frac{p\rho_1}{\rho^2}\left(1+\theta \frac{\partial \ln p^*}{\partial \theta}-\frac{\rho n_1T_1^{(0)}}{p\rho_1}\right)+\frac{\pi^{d/2}}{d\Gamma\left(\frac{d}{2}\right)}\frac{n_1 T}{\rho}\Bigg[\frac{n_1}{2}
\chi_{11}^{(0)}\sigma_{1}^d \left(1+\alpha_{11}\right)\left(\tau_1+\theta\Delta_{\theta,1}\right)\nonumber\\
& & +
n_2\mu_{12}\chi_{12}^{(0)}\sigma_{12}^d \left(1+\alpha_{12}\right)\left(\tau_2+\theta\Delta_{\theta,2}\right)\Bigg]\Bigg\},
\eeqa
where $\nu_D$ is
\beq
\label{5.21.5}
\nu_D=\nu_{11}-\nu_{12}=\frac{2\pi^{(d-1)/2}}{d\Gamma\left(\frac{d}{2}\right)}n \sigma_{12}^{d-1}\chi_{12}^{(0)}
v_0\left(1+\alpha_{12}\right)\left(\frac{\beta_1+\beta_2}{\beta_1\beta_2}\right)^{1/2}\left(x_1\mu_{12}+x_2\mu_{21}\right).
\eeq

We consider now the kinetic contribution $\eta_\text{k}$ to the shear viscosity coefficient $\eta$. The kinetic coefficient $\eta_\text{k}=\eta_1^\text{k}+\eta_2^\text{k}$, where the partial contributions $\eta_i^\text{k}$ $(i=1,2)$ obey the set of equations
\beq
\label{5.24}
\sum_{j=1}^2\left(\tau_{ij}+2\gamma_i\delta_{ij}\right)\eta_j^{\text{k}}=n_iT^{(0)}_i+\frac{\rho_iT\pi^{d/2}}
{d(d+2)\Gamma\left(\frac{d}{2}\right)}\sum_{j=1}^2n_j\mu_{ji}
\sigma^d_{ij}\chi_{ij}^{(0)}\left(1+\alpha_{ij}\right)\Bigg[\mu_{ji}\left(3\alpha_{ij}-1\right)
\left(\frac{\tau_i}{m_i}+\frac{\tau_j}{m_j}\right)
-4\frac{\tau_i-\tau_j}{m_i+m_j}\Bigg],
\eeq
where the collision frequencies $\tau_{ij}$ are defined as
\beq
\label{5.25}
\tau_{ii}=-\frac{1}{(d-1)(d+2)}\frac{1}{n_i T_i^{(0)2}}\Bigg(\int\dd\mathbf{v}R_{i,\lambda\beta}
J_{ii}^{(0)}\left[f_i^{(0)},f_{i\text{M}}
R_{i,\lambda\beta}\right]+\sum_{j=1}^2\int\dd\mathbf{v}R_{i,\lambda\beta}J_{ij}^{(0)}
\left[f_{i,\text{M}}R_{i,\lambda\beta},f_j^{(0)}\right]\Bigg),
\eeq
\beq
\label{5.26}
\tau_{ij}=-\frac{1}{(d-1)(d+2)}\frac{1}{n_j T_j^{(0)2}}\int\dd\mathbf{v}R_{i,\lambda\beta}J_{ij}^{(0)}
\left[f_i^{(0)},f_{j,\text{M}}R_{j,\lambda\beta}\right], \quad (i\neq j).
\eeq
Upon deriving Eq.\ \eqref{5.24} use has been made of the result \cite{GHD07}
\beq
\label{5.26.1}
\int \dd\mathbf{v}\; R_{i,\lambda\beta}\mathcal{K}_{ij,\lambda}\left[\frac{\partial f_j^{(0)}}{\partial V_\beta}\right]=
-\frac{\pi^{d/2}(d-1)}
{d\Gamma\left(\frac{d}{2}\right)}\rho_i n_j T \mu_{ji}
\sigma^d_{ij}\chi_{ij}^{(0)}\left(1+\alpha_{ij}\right)\Bigg[\mu_{ji}\left(3\alpha_{ij}-1\right)
\left(\frac{\tau_i}{m_i}+\frac{\tau_j}{m_j}\right)
-4\frac{\tau_i-\tau_j}{m_i+m_j}\Bigg].
\eeq

Explicit expressions of the collision frequencies $\tau_{ii}$ and $\tau_{ij}$ can be obtained by considering the Maxwellian approximation \eqref{5.18} to $f_i^{(0)}$. The results are \cite{GHD07}
\beqa
\label{5.27}
\tau_{ii}&=&\frac{2\pi^{(d-1)/2}}{d(d+2)\Gamma\left(\frac{d}{2}\right)}v_0\Bigg\{ n_i\sigma_i^{d-1}\chi_{ii}^{(0)}\left(2\beta_i\right)^{-1/2}\left(3+2d-3\alpha_{ii}\right)\left(1+\alpha_{ii}\right)
+2 n_j\chi_{ij}^{(0)}\sigma_{ij}^{d-1}\mu_{ji}\left(1+\alpha_{ij}\right)\beta_i^{3/2}\beta_j^{-1/2}\nonumber\\
& & \times \Bigg[\left(d+3\right)
\beta_{ij}\beta_i^{-2}\left(\beta_i+\beta_j\right)^{-1/2}+\frac{3+2d-3\alpha_{ij}}{2}
\mu_{ji}\beta_i^{-2}\left(\beta_i+\beta_j\right)^{1/2}+\frac{2d(d-1)-4}{2(d-1)}
\beta_i^{-1}\left(\beta_i+\beta_j\right)^{-1/2}\Bigg] \Bigg\},
\nonumber\\
\eeqa
\beqa
\label{5.28}
\tau_{ij}&=&\frac{4\pi^{(d-1)/2}}{d(d+2)\Gamma\left(\frac{d}{2}\right)}v_0n_i\chi_{ij}^{(0)}\sigma_{ij}^{d-1}\mu_{ij}\beta_j^{3/2}\beta_i^{-1/2}
\left(1+\alpha_{ij}\right)\Bigg[\left(d+3\right)\beta_{ij}\beta_j^{-2}\left(\beta_i+\beta_j\right)^{-1/2}
+\frac{3+2d-3\alpha_{ij}}{2}\nonumber\\
& & \times \mu_{ji}\beta_j^{-2}\left(\beta_i+\beta_j\right)^{1/2}-\frac{2d(d+1)-4}{2(d-1)}\beta_j^{-1}\left(\beta_i+\beta_j\right)^{-1/2}\Bigg],
\eeqa
where $\beta_{ij}=\mu_{ij}\beta_j-\mu_{ji}\beta_i$ and $i \neq j$.

Finally, the first-order contributions $T_i^{(1)}$ to the partial temperatures are defined as $T_i^{(1)}=\varpi_i \nabla\cdot \mathbf{U}$. The set of algebraic equations defining the coefficients $\varpi_i$ are given by
\beqa
\label{5.31}
& &\sum_{j=1}^2\Big[\omega_{ij}+2\gamma_j x_j\left(\tau_i+\theta \Delta_{\theta,i}\right)-2\gamma_i\delta_{ij}+\left(T_i^{(0)}+T\theta \Delta_{\theta,i}\right)\xi_j\Big]\varpi_j
=\frac{2}{d}T_i^{(0)}-\left(\zeta^{(1,0)}+\frac{2}{d}p^*\right)
\left(T_i^{(0)}+T\theta \Delta_{\theta,i}\right)\nonumber\\
& &+T\sum_{j=1}^2 \Bigg\{\frac{\pi^{d/2}}{d^2\Gamma\left(\frac{d}{2}\right)}n_j\sigma_{ij}^d \chi_{ij}^{(0)}m_{ij}(1+\al_{ij})\left[3\mu_{ji}\left(1+\alpha_{ij}\right)\left(\frac{\tau_i}{m_i}+\frac{\tau_j}{m_j}\right)
-4\frac{\tau_i}{m_i}\right]-n_j\left(\frac{\partial \lambda_1}{\partial n_j}\Delta_{\lambda_1,i}+\frac{\partial \phi}{\partial n_j}\Delta_{\phi,i}\right)\Bigg\},
\nonumber\\
\eeqa
where $\zeta^{(1,0)}$ is defined by Eq.\ \eqref{b8} and the collision frequencies $\omega_{ij}$ are
\beq
\label{5.32}
\omega_{ii}=\frac{1}{dn_iT_i^{(0)}}\Bigg(\sum_{j=1}^2\int\dd\mathbf{v}m_iV^2J_{ij}^{(0)}\left[f_{i,\text{M}}W_i,f_j^{(0)}\right]
+\int\dd\mathbf{v}m_iV^2J_{ii}^{(0)}\left[f_i^{(0)},f_{i\text{M}}W_i\right]\Bigg) ,
\eeq
\beq
\label{5.33}
\omega_{ij}=\frac{1}{dn_iT_j^{(0)}}\int\dd\mathbf{v}m_iV^2J_{ij}^{(0)}\left[f_i^{(0)},f_{j,\text{M}}W_j\right], \quad (i\neq j).
\eeq
Upon deriving Eq.\ \eqref{5.31}, we have accounted for that $\sum_j n_j \partial_{n_j}x_1=0$ and use has been made of the result \cite{GGG19b}
\beq
\label{5.33.1}
\int \dd\mathbf{v} m_i V^2 \mathcal{K}_{ij,\lambda}\left[\frac{\partial f_j^{(0)}}{\partial V_\lambda}\right]=-
\frac{\pi^{d/2}}{\Gamma\left(\frac{d}{2}\right)}\chi_{ij}^{(0)}n_i n_j \sigma_{ij}^d (1+\al_{ij})\Bigg[3\mu_{ji}(1+\al_{ij})\left(\frac{T_i^{(0)}}{m_i}+\frac{T_j^{(0)}}{m_i}\right)-4\frac{T_i^{(0)}}{m_i}\Bigg].
\eeq
Moreover, in the Maxwellian approximation \eqref{5.18}, the collision frequencies $\omega_{ii}$ and $\omega_{ij}$ read \cite{GGG19}
\beqa
\label{5.34}
\omega_{ii}&=&-\frac{\pi^{(d-1)/2}}{2dT_i^{(0)}\Gamma\left(\frac{d}{2}\right)}v_0^3\Bigg\{\frac{3}{\sqrt{2}}
n_i\sigma_i^{d-1}m_i \chi_{ii}^{(0)}\beta_i^{-3/2}
\left(1-\alpha_{ii}^2\right)-n_j m_{ij} \sigma_{ij}^{d-1}\chi_{ij}^{(0)}\left(1+\alpha_{ij}\right)\nonumber\\
& & \times \left(\beta_i+\beta_j\right)^{-1/2}\beta_i^{-3/2}\beta_j^{-1/2}
\Big[3\mu_{ji}\left(1+\alpha_{ij}\right)\left(\beta_i+\beta_j\right)-2\left(2\beta_i+3\beta_j\right)\Big]\Bigg\},
\eeqa
\beq
\label{5.35}
\omega_{ij}=\frac{\pi^{(d-1)/2}}{2dT_j^{(0)}\Gamma\left(\frac{d}{2}\right)}v_0^3n_jm_{ij}\sigma_{ij}^{d-1}\chi_{ij}^{(0)}
\left(1+\alpha_{ij}\right)\left(\beta_i+\beta_j\right)^{-1/2}\beta_i^{-1/2}\beta_j^{-3/2}\Big[3\mu_{ji}\left(1+\alpha_{ij}\right)
\left(\beta_i+\beta_j\right)-2\beta_j\Big].
\eeq
In Eqs.\ \eqref{5.34}--\eqref{5.35}, it is understood that $i\neq j$.

\end{widetext}


%

\end{document}